\documentclass[acmtog,timestamp,]{acmart}

\acmSubmissionID{279}

\usepackage{amsmath,amsthm}
\usepackage{amssymb}
\usepackage{graphicx}
\usepackage{textcomp}
\usepackage{wrapfig}
\usepackage{subfig}
\usepackage{color}
\usepackage{xspace}
\usepackage{overpic}
\usepackage{subfig}
\usepackage{enumitem}
\usepackage{booktabs}
\usepackage{algorithm2e}
\usepackage{hyperref}
\usepackage{comment}
\usepackage{svg}
\usepackage{multirow}

\definecolor{blue}{rgb}{0,0,1}
\definecolor{red}{rgb}{1,0,0}
\definecolor{green}{rgb}{0,.5,0}
\definecolor{orange}{rgb}{0.75, 0.4, 0}
\definecolor{teal}{rgb}{0.0, 0.4, 0.4}
\definecolor{purple}{rgb}{0.65,0,0.65}

\newcommand{\xy}[1]{{\color{black}\textbf{}#1}\normalfont}

\newcommand{\change}[1]{{\color{black}\textbf{}#1}\normalfont}
\newcommand{\revision}[1]{{\color{black}\textbf{}#1}\normalfont}

\newcommand{\etal}{et al.}
\renewcommand{\paragraph}[1]{\textbf{#1}}

\setcopyright{acmcopyright}\acmJournal{TOG}
\acmYear{2022}\acmVolume{41}\acmNumber{4}\acmArticle{131}\acmMonth{7} \acmDOI{10.1145/3528223.3530088}

\begin{document}

\title{Photo-to-Shape Material Transfer for Diverse Structures}

\author{Ruizhen Hu}
\email{ruizhen.hu@gmail.com}
\affiliation{
 \institution{Shenzhen University}
 \country{China}
}

\author{Xiangyu Su}
\email{xiangyv.su@gmail.com}
\affiliation{
 \institution{Shenzhen University}
 \country{China}
}

\author{Xiangkai Chen}
\email{cxk19971105@gmail.com}
\affiliation{
 \institution{Shenzhen University}
 \country{China}
}

\author{Oliver van Kaick}
\email{ovankaic@gmail.com}
\affiliation{
 \institution{Carleton University}
 \country{Canada}
}

\author{Hui Huang}
\authornote{Corresponding author: Hui Huang (hhzhiyan@gmail.com).}
\email{hhzhiyan@gmail.com}
\affiliation{%
 \institution{Shenzhen University}
 \country{China}
}

\begin{teaserfigure}
    \centering
    \includegraphics[width=\linewidth]{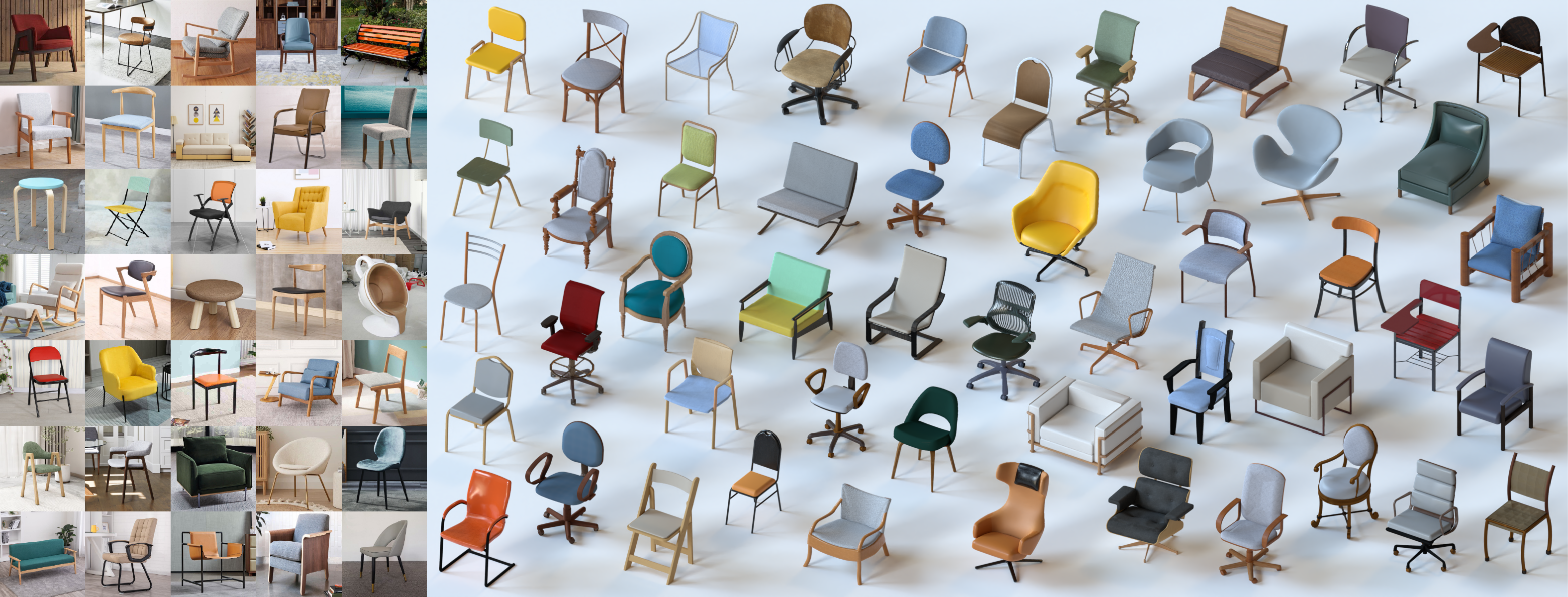}
    \caption{\revision{A sample of PhotoShapes (shapes with realistic relightable materials) created with our material assignment method by transferring materials from photo exemplars to 3D shapes with diverse structures, where the input exemplars are the photos in-the-wild on the left.}}
\label{fig:teaser}
\end{teaserfigure}

\begin{abstract}
We introduce a method for assigning photorealistic relightable materials to 3D shapes in an automatic manner. Our method takes as input a photo exemplar of a real object and a 3D object with segmentation, and uses the exemplar to guide the  assignment of materials to the parts of the shape, so that the appearance of the resulting shape is as similar as possible to the exemplar. To accomplish this goal, our method combines an \textit{image translation neural network} with a \textit{material assignment neural network}. The image translation network translates the \xy{color} from the exemplar to a projection of the 3D shape and the part segmentation from the projection to the exemplar. Then, the material prediction network assigns materials from a collection of realistic materials to the projected parts, based on the translated images and perceptual similarity of the materials. 

%based on the translated images, to maximize the similarity of the result to the input exemplar. 

One key idea of our method is to use the translation network to establish a correspondence between the exemplar and shape projection, which allows us to transfer materials between objects with \textit{diverse structures}. Another key idea of our method is to use the two pairs of (\xy{color}, segmentation) images provided by the image translation to guide the material assignment, which enables us to ensure the \textit{consistency} in the assignment. 
%We demonstrate with qualitative and quantitative evaluations 
We demonstrate that our method allows us to assign materials to shapes so that their appearances better resemble the input exemplars, improving the quality of the results over the state-of-the-art method, and allowing us to automatically create thousands of shapes with high-quality photorealistic materials.
\revision{Code and data for this paper are available at https://github.com/XiangyuSu611/TMT.}

\end{abstract}




\maketitle

\section{Introduction}

% Motivation for the problem
Most applications of computer graphics require 3D shapes with materials, since the geometry of 3D shapes alone does not fully convey the appearance of an object to a human. Shapes with materials are important for rendering realistic 3D scenes, which are relevant in simulations, games, and AR/MR/VR. Materials can include colors and textures, but high-quality realistic materials that can be relit are usually represented as reflectance functions such as BRDFs or SVBRDFs. Moreover, manually assigning realistic materials to 3D meshes can be quite difficult, since the task involves 
%splitting the mesh into patches, possibly parameterizing the patches for 2D texture mapping, and experimentally selecting material parameters for the different patches according to a desired design
experimentally selecting material parameters for patches of a 3D model according to a desired design, possibly also requiring the definition of texture coordinates. Thus, more automated ways of assigning materials to 3D shapes are necessary for mass content creation and customization.

%\ov{possibly also requiring the definition of texture coordinates.} \ov{[Do we assign texture coordinates? If not, we can remove this last sentence. Otherwise, we should also describe it in the method.]} \rh{[To estimate the UV maps we use Blender’s “Smart UV projection” algorithm as in PhotoShape.]} \ov{Added corresponding text to the method section}

% Alternative approaches other than material assignment
%\ov{[If the intro is too long, we cut the first part of this paragraph and shorten the rest]}
One option to circumvent the problem of material assignment to shapes altogether is the reuse of existing shapes. However, the shapes in many datasets such as ShapeNet and PartNet only contain basic textures but no realistic materials. In addition, shape repositories are not guaranteed to contain the shapes that the user requires; the user may need to apply materials to a specific 3D model. On the other hand, recently there have been great advances in shape reconstruction with the use of deep learning. Methods have been introduced to reconstruct shapes from different input types such as point clouds~\cite{park19,chibane20ndf}, photographs~\cite{ji17surfacenet,han21}, and sketches~\cite{wu17marrnet,delanoy18sketch}.
%with different representations such as volumes~\cite{} and implicit functions~\cite{park19,chen19}. 
Thus, this is a promising direction for the automated creation of shapes. However, in the majority of these methods, materials are not accounted for in the learning and reconstruction, and thus the generated shapes possess only their geometry.

% Introduction to the problem we solve and goals, without too many details on our method yet
In this paper, we introduce a method for automatically assigning realistic materials to 3D shapes to alleviate this gap in the automated creation of 3D shapes. Our setting is similar to that of previous work~\cite{park2018photoshape}, where our method uses photos of real objects as exemplars that guide the assignment of materials to 3D shapes, generating collections of \textit{PhotoShapes} (Figure~\ref{fig:teaser}).
%However, differently from the previous work, users can apply our method to create a PhotoShape from a specific 3D shape based on an exemplar photo, where both the shape and exemplar are not part of the datasets that we use for learning.} \ov{[Maybe this is too strong?]} 
Note that, PhotoShapes, as defined by Park \etal~\shortcite{park2018photoshape}, are not simply textured shapes, but shapes with realistic materials that can be relit, given that they are represented as reflectance functions possibly with additional texture information. As a consequence, the creation of PhotoShapes is also a different problem than simply transferring the texture of a reference photo to a 3D shape~ \cite{wang2016unsupervised}.
%,oechsle2019texture}. 

% Challenges of this problem
The creation of PhotoShapes has several challenges of its own. First, in texture transfer, the texture of the input exemplar can be directly transferred to the 3D model after the removal of illumination shading and projection distortions. On the other hand, the assignment of realistic materials requires a mapping from the material in the photograph to a database of reflectance functions, since the photograph itself does not contain such information. The material database needs to be sufficiently diverse. Moreover, the extraction of material information from the exemplar and transfer to the 3D model requires a mapping from the geometry and structure of the object in the 2D exemplar to the geometry and structure of the 3D shape, which can be a difficult matching problem.

%\ov{[I moved the new references to the related work section. I think since they don't solve exactly the same problem, they can be discussed there.]}

% Previous work and its limitations
Previous work has studied these problems and introduced methods for shape material assignment and shape reconstruction with materials. As representative examples of recent work in this area, Wang \etal~ \shortcite{wang2016unsupervised} transfer the texture from an exemplar image to a collection of shapes with an alignment method. The notable work of Park \etal~\shortcite{park2018photoshape} uses collections of shapes, materials, and photos to create thousands of PhotoShapes, where their method uses a CNN to predict material properties for shape segments. Oechsle \etal~\shortcite{oechsle2019texture} learn a spatial function that represents texture information and can be used to assign textures to 3D models.
%, while Hu \etal~\shortcite{hu21pcloud} reconstruct a textured 3D point cloud from a single RGB image. 
However, most of these methods either do not assign realistic materials or assume that a good alignment exists between the object in the exemplar and the 3D shape, and thus use local correspondence approaches to compute the part mapping.
Specifically, Park \etal~\shortcite{park2018photoshape} rely heavily on a reliable matching between photos and shapes, which is usually hard to obtain by searching example photos for a given 3D shape.

%\xy{reviewer asked to cite: 
	
%\cite{jain2012material} is the first work that introduces the task of assign material to 3D shapes without material. It gives material assignment suggestion and users can interactively change the material.

%\cite{nguyen20123d} is the first work that transfer the material from a specific photo, although it focuses more on 3D scenes and the method is more like a global optimization without 2D-3D alignment.}

\begin{figure*}[!t]
    \centering
    \includegraphics[width=0.99\textwidth]{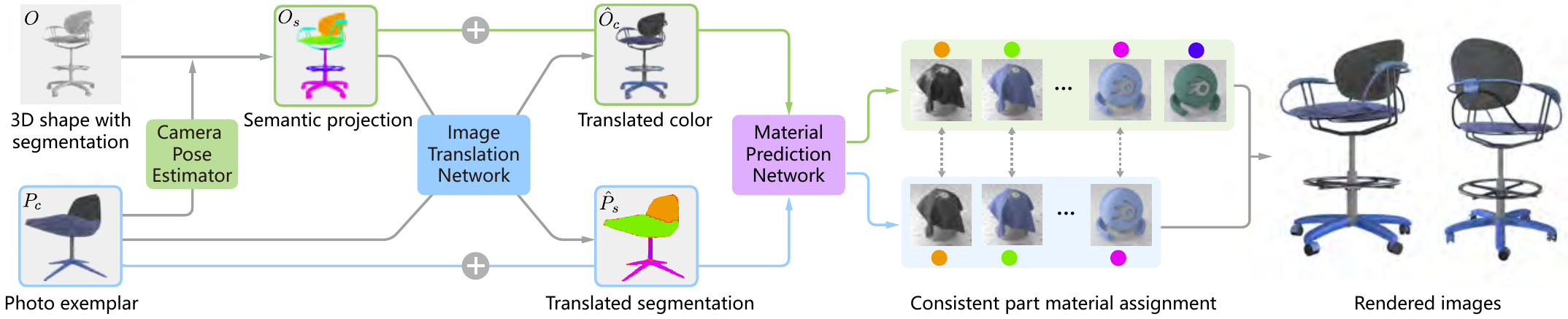}
\caption{Overview of our method for assigning photorealistic relightable
    materials to 3D shapes based on photo exemplars. Given a 3D shape
    with segmentation and photo exemplar as input, we first project the
    3D shape from a similar view as the exemplar. Next, an image
    translation network translates the \xy{color} from the exemplar to the
    projection and the part segmentation from the projection to the
    exemplar. Then, a material prediction network assigns materials to
    the projected parts based on the translated images. Note that the
    joint translation of segmentation and \xy{color} enables us to better
    ensure the consistency in the material assignment step. The result
of the method is a 3D shape that can be rendered from different
viewpoints with realistic materials.}
\label{fig:overview}
\end{figure*}

% Why our work solves the limitations of previous work and more details
% about our method
In our paper, given a 3D shape with part segmentation and a photo exemplar of a shape from the same object category, we address these limitations so that the shape and object in the photo can have very different geometric structure. One key idea of our method is that we use a state-of-the-art image translation network to establish a mapping between the parts of the two objects, which provides a more robust correspondence between the objects for material transfer. Specifically, the translation network transfers the \xy{color} from the exemplar to a projection of the 3D shape, and the part segmentation from the projection to the exemplar (Figure~\ref{fig:overview}). The correspondence is then used by a material assignment network to assign materials to the parts of the 3D object, so that the materials are similar to those in the corresponding parts of the exemplar. Another innovation over previous work is that our material assignment network accomplishes this step by considering the perceptual similarity of the materials. In addition, since the image translation results in two (\xy{color}, segmentation) pairs, we use the two image pairs to further ensure the consistency of the material assignment. 
%\ov{Our method is trained with collections of photos, shapes, and materials, but can be applied to photos and shapes outside of the collection.}
%Finally, we also introduce a more complete material library for mapping from exemplar to reflectance function \ov{[correct?]}\rh{[Their dataset has 453 materials, we have 562, but we deleted some of very similar materials in their dataset and add new ones with more diversity.]}. 

\begin{figure*}[!t]
	\centering
	\includegraphics[width=\linewidth]{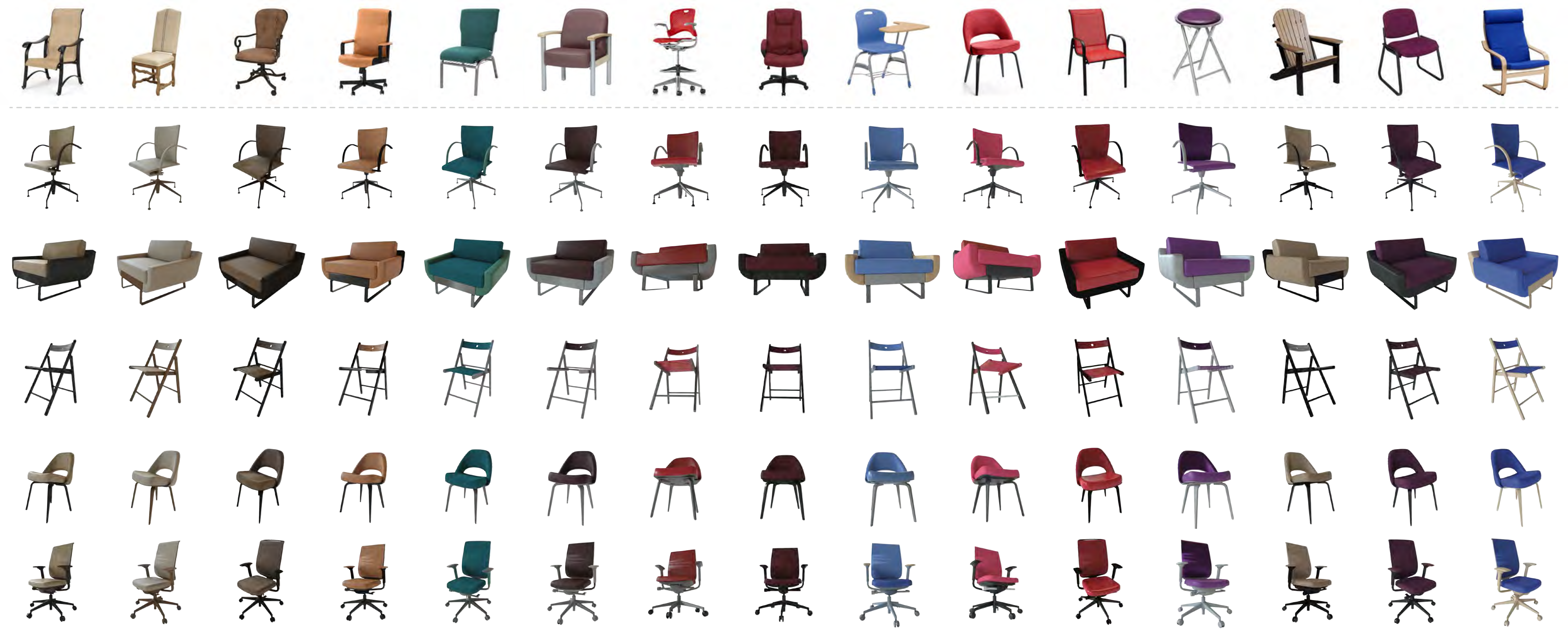}
	\caption{
	%, displaying a variety of shape structures and materials. 
Results of our method for transferring materials from exemplars to 3D shapes.
        \revision{We show results generated from different combinations of photo exemplars (given in the top row) and shapes (given in the remaining rows). Note the realism of the assigned materials, the resemblance of the results to the exemplars, and the diversity of structures in the shapes and exemplars. }
	}
	\label{fig:biggallery}
\end{figure*}

Given these contributions, we are able to create a variety of PhotoShapes (Figure~\ref{fig:biggallery}). The generated PhotoShapes have materials closer in appearance to the exemplars when compared to the results of previous work. 
We demonstrate this improvement with visual and quantitative evaluations of our method, involving a comparison to baselines and the state-of-the-art method. We also show the effect of the different components of our method in the results.
%\ov{[Maybe one more sentence about our results or say that the difference is significant]} \rh{[Yes, could also mention the comparisons to other baseline after we complete the result section to be more clear how to describe the improvement. Another improvement is that we consider perceptual similarity between materials during material prediction, and use triplet network to find a better embedded feature space to reflect such similarity thus get more accurate material classification results and smaller perceptual distance than previous method.]}

% Summary of contributions
In summary, our contributions include the introduction of:
\begin{itemize}
\item An image translation network for establishing correspondences between diverse structures in shapes and photographs, and transferring \xy{color} and segmentation information;
\item A material prediction network that assigns photorealistic materials to shape parts based on the result of image translation and a learned feature space of material perceptual similarity; 
\item A material transfer method combining the image translation and material prediction networks with a material assignment consistency criterion. 
%\item \ov{[If we plan to make this available]} A large dataset of shapes with photorealistic relightable materials containing XX models. \rh{[We don't have the exact number since we haven't check all the results we generated and how good they are in general, so maybe delete this one. ]}
\end{itemize}

%\ov{[Should we add this?]} \rh{[Could add a list to be more clear.]}

%Similarly to Park \etal~\shortcite{park2018photoshape}, our goal is to automatically produce thousands of \textit{PhotoShapes} (photorealistic relightable 3D shapes) to increase the availability of such models, which are not commonly available.

\section{Related Work}

%\ov{[Removed citation of Gao \etal and complemented missing parts.]}

Appearance modeling, especially material capture and representation~\cite{guarnera16brdf}, has received much attention in computer graphics and vision. Related problems include the development of models for establishing material similarity~\cite{lagunas19matsim} or for editing materials~\cite{serrano16matedit,schmidt16matedit}, and the introduction of large datasets of materials extracted from photographs \cite{bell13opensurf,bell15matrec}. Since our goal is to produce realistic textured materials for 3D shapes, we focus our discussion of previous work on methods related to appearance transfer onto 3D shapes. We also discuss research on image translation which is related to our solution approach.

\textit{Appearance transfer onto 3D shapes.}
The line of research most related to our work aims to transfer a combination of color, material properties, or textures from a source such as a material database, a single image, or a collection of images, to target 3D models. 
\change{Jain \etal~\shortcite{jain2012material} introduced a method to automatically assign materials to 3D objects based on correlations learned from a dataset of 3D shapes with materials. Although the user can select different material suggestions provided by the system, the user cannot directly specify the desired materials, e.g., with an exemplar. Nguyen \etal~\shortcite{nguyen20123d} introduced the first method to transfer material from an image or video to 3D geometry. However, the method focuses on transfer to 3D scenes composed of multiple objects, and uses a global optimization without 2D-3D alignment.}

Moreover, Wang \etal~\shortcite{wang2016unsupervised} transfer the texture from an exemplar input image to a collection of 3D shapes. The method rectifies the texture in the input image and then transfers the texture to a 3D shape that is geometrically similar to the object in the input image, further applying the texture to other models in the collection. 
\revision{Rematas et al.~\shortcite{rematas2016novel} align a target 3D shape to a reference 2D photo and extract per-segment materials.}
Liu \etal~\shortcite{liu17color} color a 3D model based on an example photograph by establishing point correspondences between the image and 3D model, which are then used to transfer color information from the image to the model. Moreover, Zhu \etal~\shortcite{zhu18scene} colorize 3D furniture models and indoor scenes according to an image exemplar, based on transferring color information across image segmentations. Huang et al.~\shortcite{huang18proxy} facilitate the transfer of a detailed texture from a photograph to a 3D model by aligning a simple proxy to the object in the photograph, which allows the method to extract the geometric detail of the object and a reflectance texture from the photograph. The extracted information can then be applied to new models.

With the advent of deep learning, Park \etal~\shortcite{park2018photoshape} use shape collections, material collections, and photo collections to assign materials to 3D shapes with a CNN that predicts material properties for shape segments. Differently from most related work, their method assigns materials represented as SVBRDFs, producing photorealistic relightable 3D shapes (PhotoShapes), rather than models with only color or texture information. Raj \etal~\shortcite{raj19texturegen} also use collections of 3D models and images to generate textured models. However, they perform the transfer in the image domain by rendering a model from a set of viewpoints and assigning textures to the resulting images, which are then fused together into a textured 3D model. Mir \etal~\shortcite{mir20clothing} specialize this type of method to transfer textures from images of clothing to 3D garment models. 
%Recently, in unpublished work, Gao \etal~\cite{gao20tmnet} introduce a shape representation based on deformable, textured boxes, which can be used to automatically assign textures to models or interpolate textures and shapes in a joint latent space. \rh{[Maybe remove the citation of \cite{gao20tmnet}?]}
%The method learns a mapping from 2D images to 3D models according to
%the silhouette of the garment's 3D geometry. 

\change{Since we need to establish a correspondence between the exemplar image and 3D shape before appearance transfer, the material transfer pipeline can be naturally decomposed into two subtasks: 2D-3D alignment and information transfer. Thus, methods that address the general problem of 2D-3D alignment are also relevant to our work. Specifically, Rematas \etal~\shortcite{rematas2016novel} align renderings of segmented shapes to images through a coarse-to-fine refinement method. 
%The method first renders 3D shapes from a dataset from different views, and retrieves a rendering similar to target image, so that a correspondence can be established. 
Su \etal~\shortcite{su2014estimating} introduce a method that extracts silhouettes from renderings and an input image to establish a 2D-3D correspondence.
%based method to find a correspondence between 2D and 3D data. The method first renders shapes from the camera pose estimated for the image, and then applies a continuous time warping~\cite{munich1999continuous} to build a dense correspondence between silhouette curves which guide the alignment between the image and rendering.
Huang \etal~\shortcite{huang2015single} use a method based on matching of patches to establish a 2D-3D correspondence. 
%They first render shapes from different views, extract patches from renderings and images, construct patch graphs for similar patches. By clustering patch graphs, patch correspondence can be found. And based on patch correspondence, target image can be segmented into several parts, so part correspondence between 2D and 3D data can be established.

}

%\xy{[More works on 2D-3D alignment]

%It’s obvious that correspondence between exemplar image and 3D shape must be established before appearance information transfer, so the whole pipeline can be decomposed into two different subtasks, 2D-3D alignment establishing and information transfer. Previous research provides us different ways to align images and shapes. \cite{rematas2016novel} align part segmented 3D shape to image through Coarse to fine method. They first render 3D shapes from different views, and retrieve a 3D shape rendering similar to target image from 3D shape datasets, and then refine the view of shape continuously, use simulated annealing to refine individual parts. After that, each part in 3D shape can find its corresponding region in target image. \cite{su2014estimating} proposed a silhouette-based methods to find correspondence between 2D and 3D data. They first render shapes from estimated camera pose, and apply \cite{munich1999continuous} to build dense correspondence between silhouette curves, then employ Laplacian deformation to align shape rendering and target image.  \cite{huang2015single} use path-based method to get correspondence. They first render shapes from different views, extract patches from renderings and images, construct patch graphs for similar patches. By clustering patch graphs, patch correspondence can be found. And based on patch correspondence, target image can be segmented into several parts, so part correspondence between 2D and 3D data can be established.}

Departing from the idea of an information transfer pipeline as introduced by the works discussed above, Oechsle \etal~\shortcite{oechsle2019texture} propose to represent texture information as a function in 3D space, similar to an implicit occupancy or signed distance function for representing a 3D shape~\cite{park19,chen19}. The function, coined a ``texture field'', is learned with a set of neural networks, where the generation of the texture color for a 3D point is conditioned on an image and a shape. On the other hand, Hu \etal~\shortcite{hu21pcloud} reconstruct a textured 3D point cloud from an input image, since points clouds are a more lightweight representation than volumes, meshes, or implicit functions. However, the resulting model and texture produced by the method are fixed by the input. 

% Zhu et al.: It is easier to predict materials based on object type rather than the color the objects are painted with

\textit{Discussion.}
The work of Park \etal~\shortcite{park2018photoshape} is the most related to our work. Similarly to their work, our goal is to produce realistic PhotoShapes. However, differently from their solution, our method does not require an existing match between photos and shapes in the database. Instead, our method is able to transfer information from a photo exemplar with a significantly different geometric structure from the given 3D shape by relying on a matching of part segments obtained by image translation. This is more robust than the local alignment method of Park \etal~\shortcite{park2018photoshape} that requires matching points to be sufficiently close, as it applies a projection followed by SIFT flow. Moreover, as discussed above, other works assign mainly colors or texture information to the 3D models rather than a complete material specification, and also have similar assumptions of geometric similarity of the photo's object to the input 3D model or proxy shape. 
%Specifically, the method of Gao \etal~\shortcite{gao20tmnet}, although also based on a part analysis as our method, only assigns texture information to 3D models, and the output shape is represented as a set of deformable boxes, rather than the input model with transferred textures. 
Furthermore, the texture field method, although interesting as an alternative to direct material transfer, requires the input model to be retrained for each input exemplar and does not assign 2D texture coordinates to the input 3D shape, resulting in textures that have mainly uniform colors.
%\rh{[can also remove the comparison to\cite{gao20tmnet} here?]}

\textit{Image translation.}
Exemplar-based image translation can be characterized as a form of conditional image synthesis where an input structure (segmentation mask, edge map, or set of pose keypoints) is converted into an image with the style of an input exemplar~\cite{isola2017trans,chen17synth,huang18trans}. This form of image translation requires to map the input structure to the exemplar image, establishing a \textit{cross-domain} correspondence. In our work, we use image translation to map the texture of an input exemplar into the segmentation of a rendered image, and vice-versa, which serves as the first step of our material assignment pipeline. Specifically, we use the method of Zhang et al.~\shortcite{zhang20imagetrans}, where image translation and cross-domain correspondence are learned together with two jointly-trained neural networks, so that the solution of each task facilitates the other.

%\cite{bell13opensurf} a dataset of surfaces segmented from images and
%annotated with material and texture information. The rectified textures
%can then be used in models 
%\cite{bell15matrec} create a dataset of materials and use it to segment
%and recognize materials in new images with 3 million samples

\section{Overview}

Figure~\ref{fig:overview} shows an overview of our method. The input to our method is a 3D shape with part segmentation and a photo exemplar of an object from the same category, which may have a totally different geometric structure than the given shape. We assume that the 3D shape is given with the part segmentation. However, note that such segmentation can be obtained automatically with existing methods~\cite{kalogerakis17seg,mo2019partnet,pham19seg}. Our method starts by estimating the camera pose of the photo exemplar.
Then, we project the 3D shape based on the camera pose and transfer the part segmentation from the 3D shape to the 2D projection, resulting in a labeled image that we call a \textit{semantic projection}. 
%generate a 2D projection \rh{[should we change "2D projection" to "semantic projection"?]}of the given 3D shape, which corresponds to the same view as the exemplar and carries over the part segmentation from the model to the projection. 
% OV: I added the definition of semantic projection here

Next, our image translation network takes the photo exemplar and the semantic projection as input, and outputs two translated images. One image translates the \xy{color} from the exemplar to the projection, and the other image translates the part segmentation from the projection to the exemplar. Note that we use the term \textit{translation} in a similar manner as the related literature~\cite{zhang20imagetrans}, i.e., a transfer of \xy{color} or structure information from one image to another, not to be confused with the spatial transformation. As a result, we get two (\xy{color}, segmentation) image pairs. For each pair, we use our material prediction network to obtain a material assignment to the object parts, while at the same time we use the two image pairs to further ensure the consistency in the material assignment. Note that the use of two pairs of images is a key component of our method to ensure the assignment consistency and thus improve the quality of the result. The final part material assignment can then be used to render realistic images of the 3D shape with similar material as the photo exemplar.
%We train the two networks that compose our method with the datasets described as follows.

\revision{
\paragraph{Datasets.}
Since our method is based on learning, we use collections of shapes, photographs, and materials for training our neural networks and evaluating our results. 
Note that shape and photograph collections are category-specific, while we use the same material collection for the different object categories tested. 
We extend the material collection provided by Park \etal~\shortcite{park2018photoshape} by adding new, more distinctive materials, which results in a set of 600 materials in total. We group the materials into five categories: leathers, fabrics, woods, metals, and plastics. 
Since the collections do not directly provide samples of translated images and shapes, we use the collections to create synthetic data for training our networks, as in the work of Park \etal~\shortcite{park2018photoshape}. %, but with a more diverse material dataset and training pairs. 
We describe more details about the collections and data generation in the supplementary material.
}

\section{Material transfer method}
\label{sec:method}

In this section, we first explain the details of the two key components of our method: the networks for image translation and material prediction. Then, we explain how we combine the two networks together to perform the material transfer. 

\subsection{Image translation}
\label{sec:translation}

Given the image exemplar, we first estimate the camera pose of the image and use it to generate the semantic projection (2D projection with part segmentation) of the given 3D shape. To obtain the camera pose, we adopt the camera pose estimation network proposed by Xu et al.~\shortcite{xu2019disn}. Then, we project the 3D shape based on the camera pose and transfer the part segmentation from the 3D shape to the 2D projection to create the semantic projection. 
%\ov{[I defined the term \textit{semantic projection} here. So, we don't need to always explain that it is a projection with part information, and to keep consistency we can use this term instead of labeled image or mask. But, I also used the term \textit{segmentation} when referring specifically to the parts rather than the entire semantic projection.].} \rh{[Sounds good. Just to note that I also used \textit{mask} in the network figures.]}

Moreover, given the photo exemplar and the semantic projection, the goal of image translation is to generate both the part segmentation for the photo exemplar and the \xy{colored} image for the semantic projection so that the \xy{colored} image and semantic projection are in correspondence with each other. The two pairs of images are then used for part material prediction in the next step.  However, the key challenge here is that such images with translated \xy{color} do not naturally exist in the wild and thus we need to learn to establish a cross-domain correspondence between the exemplar and projection. To tackle this problem, we use the method of Zhang et al.~\shortcite{zhang20imagetrans}, where image translation and cross-domain correspondence are learned together with two jointly-trained neural networks.

%\rh{Network architecture and loss definition.}

\begin{figure}[!t]
	\centering
	\includegraphics[width=\linewidth]{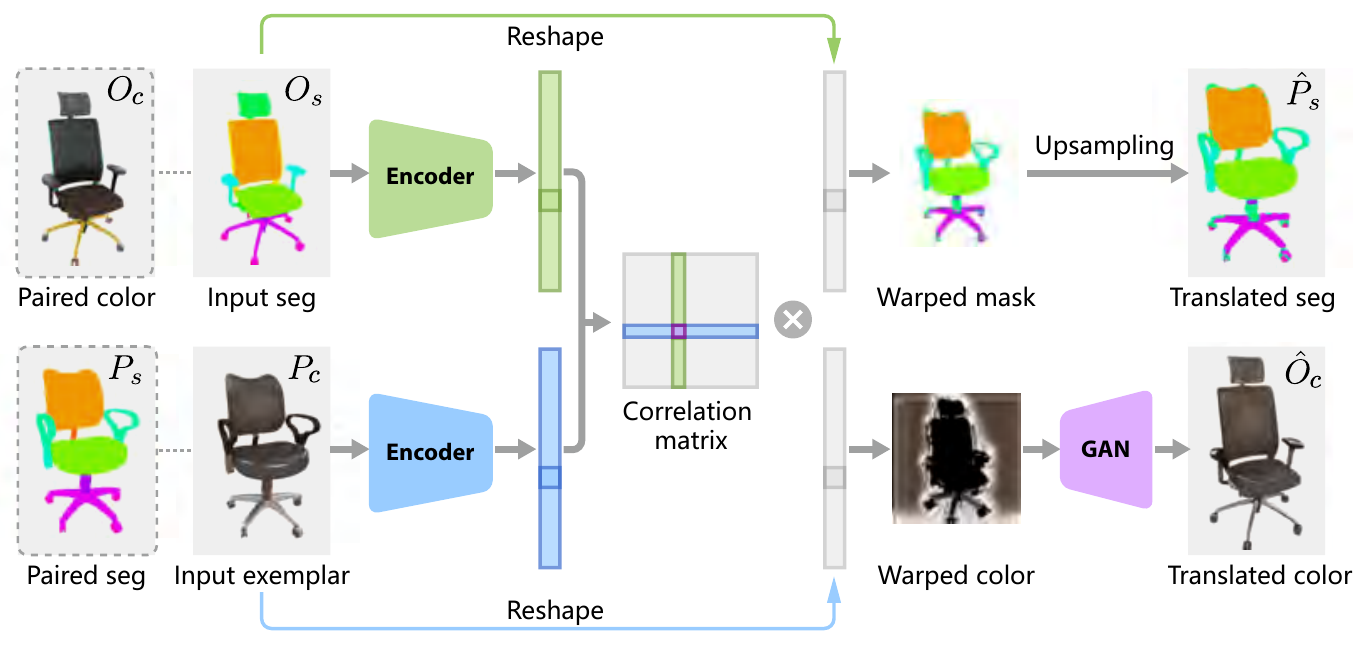}
	\caption{
            Image translation method used in our work, where the core components are based on the method of Zhang et al.~\shortcite{zhang20imagetrans}. Given the input semantic projection $O_s$ of a 3D object and the photo exemplar \xy{$P_c$}, we establish a correspondence between the images, which is then used for image warping and information transfer, resulting in translated segmentation $\hat{P}_s$ and translated \xy{color $\hat{O}_c$} . 
        %The outputs of the translation are then further improved. 
        Note that the paired \xy{color $O_c$} and segmentation mask $P_s$ are only used for training. Please refer to the text in Section~\ref{sec:translation} for more details.}
%    the two images are first embedded into a common domain, where a correspondence between the images is estimated, which is represented as correlation matrix. Next, the input images are warped according to the correspondence. The warped semantic projection is upsampled to the resolution of the input exemplar, while the appearance of the translated texture is improved with an image synthesis network.
	\label{fig:translation_net}
\end{figure}

Figure~\ref{fig:translation_net} shows a simplified diagram of the image translation method~\cite{zhang20imagetrans}. Given the  semantic projection of  the input object $O_s$ and the photo exemplar with \xy{color $P_c$}, a first network embeds the two images into a common domain where it is possible to build a dense correspondence with a correspondence layer~\cite{zhang19corresp}. The correspondence is represented by a correlation matrix. Then, the input images are warped according to the correspondence, resulting in the translation of segmentation and \xy{color}. After that, we upsample the translated segmentation to the same resolution as the input exemplar to obtain the translated segmentation $\hat{P}_s$, and apply an image synthesis GAN to the translated color to ensure that the generated image \xy{$\hat{O}_c$} looks more natural. 
\change{Note that, in our notation, $s$ and \xy{$c$} refer to the two different translation domains (segmentation vs. \xy{color}), and $O$ and $P$ refer to the different image contents (projection of 3D object vs. photo exemplar).%(segmentation, \xy{color}).
}

The main challenge of this cross-domain image translation problem is to learn the correspondence without direct supervision, since we do not have a ground truth \xy{$\hat{O}_c$} for the final output. Creating such ground truth is a difficult problem. %On the other hand,
%% as described in Section~\ref{sec:datasets}, 
%the data that we have available for supervision is 
\change{Therefore, we instead make use of images that are easier to create, including}
the \xy{colored} image \xy{$O_c$} corresponding to the input semantic projection $O_s$ and the segmentation image $P_s$ corresponding to the input \xy{colored} exemplar \xy{$P_c$}. This data is illustrated in the dashed boxes on the left of Figure~\ref{fig:translation_net}. Thus, during the training of the translation method, quadruples $(\xy{O_c}, O_s, \xy{P_c}, P_s)$ composed of two (\xy{color}, segmentation) pairs are used to guide the training.

\change{
The loss function that guides the training the image translation network is defined as:
\begin{equation}
\label{eq:trans}
     L_t  =  \xy{L_{\text{col}}} + L_{\text{seg}} +  L_{\text{reg}}, 
\end{equation}
where \xy{$L_{\text{col}}$} is the \xy{color} loss, $L_{\text{seg}}$ is the segmentation loss, and $L_{\text{reg}}$ is the regularization loss.  The first two terms ensure that the translated \xy{color} and segmentation are similar to the corresponding ground truth data and look natural, while the last term regularizes the correspondence between the input pairs from two different domains. 
%\rh{We will give the explanation of general goal of each term as follows and leave more detailed definition in the supplementary material.}

\subsubsection{\xy{Color} loss.}
%The loss $L_{\text{tex}}$ is defined on the translated \xy{color} $\hat{O}_c$. 
%First, we add two terms $L_{\text{context}}^{\text{tex}} $ and $L_{\text{perc}}$ ensure that its low level feature is similar to $P_c$ as the \xy{color} is translated from $P_c$  and its high-level feature is similar to $O_c$ as it shares the same structure with $O_c$. 
%Then,  we generate pseudo exemplar pairs and add a term $\psi_3 L_{\text{feat}}$ to encourage $\hat{O}_c$ to be similar to the  $O_c$ when using ${O}^{'}_t$ as the exemplar, which is obtained by applying random geometric distortion to $O_c$.
%Finally, to ensure that the synthesized images $\hat{O}_c$ look indistinguishable from real ones, we add the adversarial loss $L_{\text{adv}} $.
%Thus $L_{\text{tex}}$ is defined as:
The loss \xy{$L_{\text{col}}$} is mainly used to constrain the translated \xy{color} $\hat{O}_c$, and is defined as:
\begin{equation}
	\label{eq:col}
	L_{\text{col}} = \psi_1 \xy{L_{\text{context}}^{\text{col}}} + \psi_2 L_{\text{perc}}+  \psi_3 L_{\text{feat}} + \psi_4 L_{\text{adv}} ,
\end{equation}
where 
1) \xy{$L_{\text{context}}^{\text{col}}$}  is the context loss to minimize the low-level feature distance between \xy{$\hat{O}_c$} and \xy{$P_c$} for any given exemplar \xy{$P_c$} as the \xy{color} is translated from \xy{$P_c$};
2) $L_{\text{perc}}$ is the perceptual loss to minimize the high-level feature distance between \xy{$\hat{O}_c$} and \xy{$O_c$} for any given exemplar \xy{$P_c$}, as it shares the same structure with \xy{$O_c$};
3)  $L_{\text{feat}}$ is the feature matching loss to minimize the feature distance between \xy{$\hat{O}_c$} and \xy{$O_c$} when the input is a pseudo-exemplar pair $(O_s, \xy{{O}^{'}_c})$, with \xy{${O}^{'}_c$} obtained by applying random geometric distortion to \xy{$O_c$}, as the \xy{color} is kept the same during the distortion;
4) $L_{\text{adv}} $ is the adversarial loss to ensure that the synthesized images \xy{$\hat{O}_c$} look indistinguishable from real ones.

%\begin{itemize}
%\item $L_{\text{context}} $  is the context loss to minimize the low-level feature distance between $\hat{O}_c$ and $P_c$ for any given exemplar $P_c$ as the \xy{color} is translated from $P_c$ ;
%\item $L_{\text{perc}}^{\text{tex}}$ is the perceptual loss to minimize the high-level feature distance between $\hat{O}_c$ and $O_c$ for any given exemplar $P_c$, as it shares the same structure with $O_c$;
%\item $L_{\text{feat}}$ is the feature matching loss to minimize the feature distance between $\hat{O}_c$ and $O_c$ when the input is a pseudo-exemplar pair $(O_s, {O}^{'}_t)$, with ${O}^{'}_t$ obtained by applying random geometric distortion to $O_c$, as the \xy{color} keeps the same during the distortion;
%\item $L_{\text{adv}} $ is the adversarial loss to ensure that the synthesized images $\hat{O}_c$ look indistinguishable from real ones.
%\end{itemize}

\subsubsection{Segmentation loss}	
The loss $L_{\text{seg}}$  is mainly used to constrain the translated segmentation $\hat{P}_s$, and is defined as:
\begin{equation}
	\label{eq:seg}
	L_{\text{seg}} = \psi_5 L_{\text{pred}} + \psi_6 L_{\text{context}}^{\text{seg}},
\end{equation}
where
 $L_{\text{pred}}$ is the prediction loss to minimize the distance between $\hat{P}_s$ and $P_s$ for any given semantic segmentation mask $O_s$,
and $L_{\text{context}}^{\text{seg}} $  is the perceptual loss  to minimize the low-level feature distance between  $\hat{P}_s$ and $P_s$ for a given semantic segmentation mask $O_s$.
%\begin{itemize}
%	\item $L_{\text{pred}}$ is the prediction loss to minimize the distance between $\hat{P}_s$ and $P_s$ for any given semantic segmentation mask $O_s$;
%	\item $L_{\text{perc}}^{\text{seg}} $  is the perceptual loss to minimize the high-level feature distance between $\hat{P}_s$ and $P_s$ for given semantic segmentation mask $O_s$;
%\end{itemize}

\subsubsection{Regularization loss}
The loss $L_{\text{reg}}$ is used to regularize the embedding in the shared domain of both inputs, and is defined as:
\begin{equation}
	\label{eq:reg}
	L_{\text{reg}} = \psi_7 L_{\text{align}} + \psi_8 L_{\text{cycle}} ,
\end{equation}
where
$L_{\text{align}}$ is the domain alignment loss to minimize the feature distance between $O_s$ and \xy{$O_c$} after embedding in the shared feature space, 
and $L_{\text{cycle}} $ is the correspondence regularization to ensure that the final output after a cycle process is similar to the input exemplar \xy{$P_c$}.
For more details about the network architecture and loss definition, please refer to the supplementary material.
% the work of Zhang et al.~\shortcite{zhang20imagetrans}. 

}

\subsection{Material prediction}

Given a pair composed of a \xy{colored} image and its corresponding part segmentation image, the goal of the material prediction network is to obtain an optimal part material assignment for the input, similar to the method of Park et al.~\shortcite{park2018photoshape}. However, differently from their method, which treats the material prediction problem purely as a classification problem, we take the perceptual similarity between materials into consideration to assign materials that maximize the visual similarity between the material predicted for each segment and its corresponding patch in the \xy{colored} image. The motivation for using the perceptual similarity is that then the network can perform a suitable material assignment even if the material label is incorrect.
%\ov{[or substance label?]} \rh{[if the material label is incorrect, then substance label could still be correct, so it would be a hander case?]}
Thus, this step requires to learn the perceptual similarity between predicted materials and patches in the \xy{colored} image.

To learn the perceptual similarity, we first prepare material images that will be used in the learning. We render an image for each material in our dataset with the specific scene and perspective settings proposed by Havran et al.~\shortcite{havran2016perceptually}, which optimize the coverage of the BRDF in the image. %Figure~\ref{fig:material} shows the rendering of representative materials. 
Then, we compute the L2-lab distance between the images~\cite{sun2017attribute}. The reason for choosing the L2-lab distance as the perceptually-based metric is that this distance is closer to human perception than other metrics, according to a recent comparison~\cite{lagunas2019similarity}. Given a material dataset with $n$ different materials, we compute the L2-lab distance between each pair of materials to form the pairwise distance matrix $D \in \Bbb R^{n \times n}$. 
Given this material perceptual information, we train a neural network to assign materials that minimize the perceptual difference between the predicted and ground truth materials, \change{where the classification task is combined with metric learning.}

\begin{figure}[!t]
	\centering
	\includegraphics[width=\linewidth]{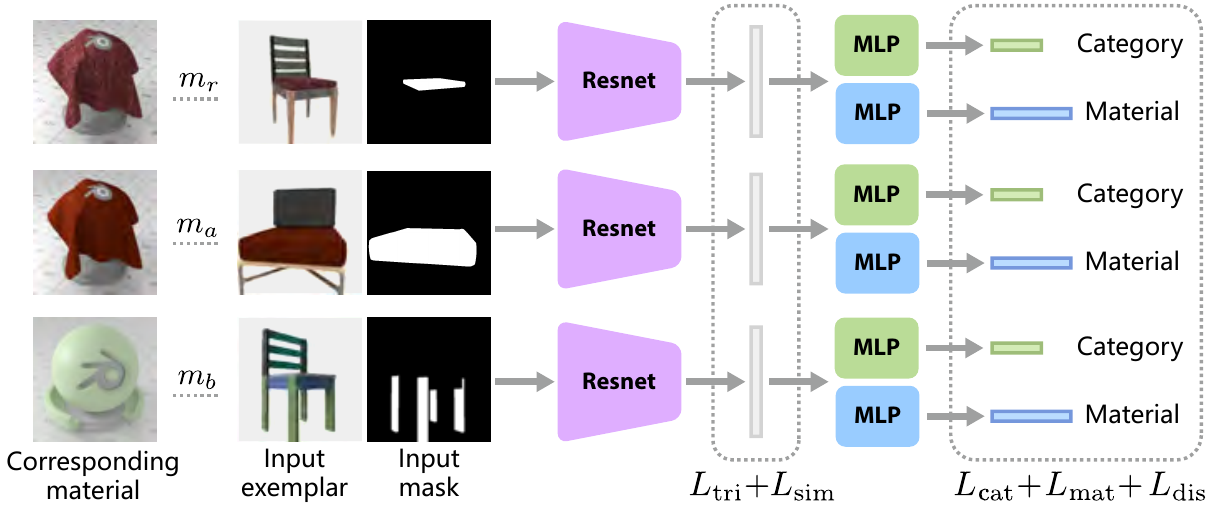}
        \caption{Material prediction network used in our work. During training, we use a triplet network composed of ResNets with shared weights (middle) to learn to embed the input images into a feature space that reflects the perceptual similarity of the materials in the images (left), and then we further learn two classifiers based on the feature space to predict the overall material category and material label (right). After training, the network can be used to predict the material category and label for a single input \xy{exemplar} and selected part. The diagram also indicates the terms of the loss function used for training each part of the network.}
        %\ov{[Rename substance in the figure? The term in the loss also has to be changed to $L_{cat}$]}}
%    \rh{[Actually, the left three input is the sampled material triplet, and the corresponding texture and mask pairs are taken as input of the triplet networks to embed them input the common feature that ensure the material similarity satisfy the requirement that the positive example should be similar to the reference than the negative one. Then two MLPs are added to predict  the substance and material labels to make the ground truth lable given in the left, at the same time minimize the perceptual distance between the predicted material and the ground truth material.]}    
%    \ov{[In the overview figure, the network takes a pair of images as input and returns the material assignments. I think we need 1-2 sentences at the beginning of Section 4.2 to explain how we use this triplet network in the way shown in the overview. The training part is clear, just need some clarification about test time.]}
%\rh{[Since the triplet network use the shared networks, so during the testing, we just need one pair of texture and mask to perform the classification.]} The diagram also indicates the terms of the loss function used for training each part of the network.}
	\label{fig:prediction_net}
\end{figure}

Figure~\ref{fig:prediction_net} shows the architecture of our material prediction network. 
We use a Resnet-34~\cite{he2016deep} pretrained on ImageNet \cite{deng2009imagenet} as our backbone network, but remove the last layer and add a FC layer instead to embed the input images into a $128$-D feature space. \revision{We also add two FC layers after that to predict the material category and label, respectively.
Inspired by the work of Lagunas et al.~\shortcite{lagunas2019similarity}, we use a triplet network~\cite{hoffer2015deep} to embed the input images into a perception-aware feature space where the material similarity can be established and then the material labels can be assigned.
For training the network and learning the feature space, we input three pairs composed of a \xy{color} image and a part mask indicating the selected part.
Once the network is trained, we input only one pair and obtain the material category and labels for the selected part.}
%other than the classification error.

The loss function for training our material prediction network is defined as follows:
\begin{equation}
	\label{eq:matpred}
	L_p = L_{\text{metric}} +  L_{\text{class}} ,
\end{equation}
where $L_{\text{metric}}$ is the metric learning loss to ensure the embedded features reflect the perceptual distance, and  $L_{\text{class}} $ is the loss defined on the final material assignment to make sure it is similar to the ground truth.

\subsubsection{Metric learning loss.} 
The metric learning loss $L_{\text{metric}}$  is adapted from the work of Lagunas et al.~\shortcite{lagunas2019similarity}, and composed of two losses  defined on material triplets, where each triplet given to the network consists of a reference image $r$ with material $m_r$, one positive example $a$ with material $m_a$, and one negative example $b$ with material $m_b$. 
The metric learning loss is defined as:
\begin{equation}
	\label{eq:matpred}
	L_{\text{metric}} =\alpha_1 L_{\text{tri}} + \alpha_2 L_{\text{sim}},
\end{equation}
where $L_{\text{tri}}$ is the triplet loss that seeks to bring similar materials $r$ and $a$ closer together in the feature space and repel dissimilar materials $b$, and $L_{\text{sim}}$ is the similarity loss that further maximizes the log-likelihood of the model choosing $a$ to be closer to $r$ than $b$.

 The triplet loss is defined as follows~\cite{lagunas2019similarity}:
\begin{equation}
	L_{\text{tri}}(r, a, b) = \frac{1}{|B^A|} \sum_{(r,a,b) \in B^A} \left[ \lVert f_r - f_a\rVert_2^2  - \lVert f_r - f_b\rVert_2^2  + \mu \right]_{+},
\end{equation}
where $[x]_{+} = \max(0,x)$, $f_x$ is the feature vector of $x$, $\mu$ is the margin which specifies how much we would like to separate the samples in the feature space, and $B^A$ is the set of triplets that take part in the training.
%\ov{[Should we change it to $m_r$ instead of $m^r$ for consistency with the notation above? Or are these single labels rather than vectors? Then, we can clarify this.]} \rh{[Right, let's use $m_r$ to be more consistent.]}

The similarity loss is defined as follows~\cite{lagunas2019similarity}:
\begin{equation}
	L_{\text{sim}}  = -\frac{1}{|B^A|} \sum_{(r,a,b) \in B^A} \log \frac{s_{ra}}{s_{ra} + s_{rb}},
\end{equation}
where $s_{ra} = \dfrac{1}{1+\lVert f_r - f_a\rVert_2^2}$ and $s_{rb} = \dfrac{1}{1+\lVert f_r - f_b\rVert_2^2}$.
\\
\\
\revision{
    To construct the training triplets $B^A$, we first generate a set of pre-sampled material triplets $A^M$. To generate $A^M$, we randomly sample a reference material $m_r$, a random sampled positive material $m_a$ which is of the same category as $m_r$, and a negative material $m_b$ which is of a different category than $m_r$. The material $m_b$ is sampled so that it has a larger perceptual distance to $m_r$ than $m_a$, i.e., $D(m_r,m_b) > D(m_r,m_a)$, but also has a distance smaller than the distances to all other materials of all different categories, i.e., $D(m_r,m_b) < D(m_r,m_x)$, where $m_x$ is any material of a category other than the category of $m_r$.} %and $m^b$.
Then  $B^A$ is defined as:
\begin{equation}
	B^A = \left\{(r,a,b) \mid (m_r, m_a, m_b) \in A^M  \wedge (r, a, b)\in B \right\},
\end{equation}
where $B$ is the current training batch. Thus, $B^A$ is the set of all triplets of images in $B$ whose corresponding material labels appear as triplets in $A^M$.

\change{
\subsubsection{Classification loss.} 
The classification loss $ L_{\text{class}}$ for training our material prediction network is defined as follows:
\begin{equation}
	\label{eq:matpred}
	 L_{\text{class}} = \alpha_3 \xy{L_{\text{cat}}} + \alpha_4 L_{\text{mat}} + \alpha_5 L_{\text{dis}},
\end{equation}
\revision{where \xy{$L_{\text{cat}}$} and $L_{\text{mat}}$ are the cross entropy losses defined by Park et al.~\shortcite{park2018photoshape} for material category and label classification, and $L_{\text{dis}}$ is the distance loss to minimize  the perceptual distance between the predicted material and the ground truth material:}
\begin{equation}
	\label{eq:dist}
    L_{\text{dis}}(m_p, m_{gt}) = m_p^{T} D_{\text{idx}(m_{gt})},
\end{equation}
where $m_p$ and $m_{gt}$ are the predicted and ground truth material labels, represented as $n$-dimensional column vectors where $n$ is the size of the material dataset. Specifically, these vectors are one-hot vectors for the ground truth labels, and probability vectors for predicted labels.
%\ov{[Are these one-hot vectors then?]} \rh{[Yes for ground truth label. But for the predicted label it would just be a |M|-D vector indicating the probability of each label. From the previous notation, then |M| will just be $n$, right? ]}
$D_i$ represents the $i$-th column of the matrix $D$, $\text{idx}(m_{gt})$ is the index of the ground truth material and thus $D_{\text{idx}(m_{gt})}$ encodes the perceptual distance of $m_{gt}$ to all other materials. %\rh{[Should we use different notations? Since $x$ and $y$ have been used in the image translation network. Maybe use $m_p$ and $m_{gt}$?]}
%\ov{[Added "column" above and changed the notation a bit for the formula to make sense and L be a scalar. Is this fine?]} \rh{[Yes. Maybe we can use $D(:,m_{gt})$ to represent the corresponding column and simplify the notation since $m_{gt}$ is a one-hot vector.]}\ov{[I thought it's better to use proper mathematical notation, even though it's a bit more complicated, as some reviewer may complain]} 
}

\begin{figure}[!t]
	\centering
	\includegraphics[width=0.98\linewidth]{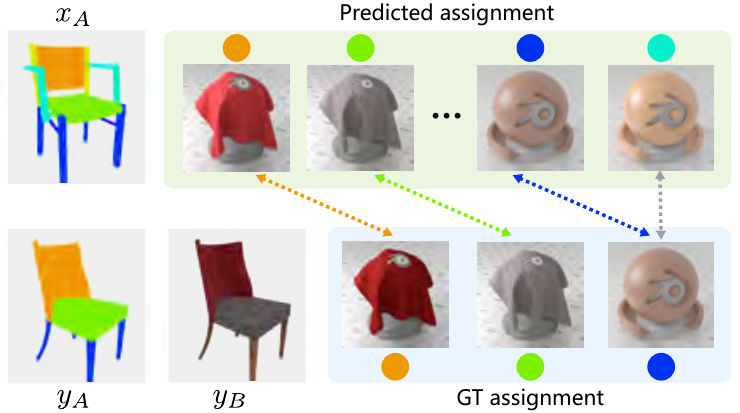}
	\caption{
        Method for constructing the ground truth part material assignment for (\xy{color}, segmentation) pairs generated by the image translation network. This ground truth is used for computing the loss of the material prediction network. When defining the ground truth of $O_s$, for corresponding parts that exist in $P_s$, the ground truth assignment is the corresponding part material indicated by the colored arrows. For parts that do not exist in $P_s$ (cyan material on the top-right), the ground truth assignment is the part material in the ground truth assignment with minimal perceptual distance to the predicted material, indicated by the gray arrow. } 
	\label{fig:assign_part}
\end{figure}

\subsection{Material transfer} 
\label{sec:mattransf}
As illustrated in Figure~\ref{fig:overview}, the entire pipeline for material transfer from the given photo exemplar to the 3D shape consists of two key steps, i.e., image translation and material prediction. 
We first pre-train the image translation and material prediction networks separately, and then fine-tune the material prediction network to provide consistent part material assignments for two (\xy{color}, segmentation) image pairs, i.e., ($\xy{P_c}, \hat{P}_s$) and ($\xy{\hat{O}_c}, O_s$). 
\change{Note that the two (\xy{color}, segmentation) image pairs have different input features, since the translated \xy{color $\hat{O}_c$} may be noisy and the translated segmentation $\hat{P}_s$ can be inaccurate. Thus, we fine-tune two versions of the material prediction network for each of the two pairs separately, and then use a consistency loss to enforce a consistent part assignment.}

\change{
Moreover, when training the material prediction network alone, the required training data with ground truth can be easily generated given the dataset we collected. However, it is challenging to create ground truth data for training the entire pipeline together, since we require corresponding translated (\xy{color}, segmentation) image pairs, but these are difficult to create due to the existence of missaligned structures in the images and unmatched semantic parts. 
 We use different methods to determine the ground truth for the two different types of translated (\xy{color}, segmentation) image pairs.}
 %We can only use partial data from those pairs to fine-tune the prediction network.

\change{
\subsubsection{Fine-tuning with translated \xy{color}.} 
%For ($\hat{O}_c, O_s$)  pairs, the \xy{color} $\hat{O}_c$ is translated from exemplar $P_c$, thus we use the ground truth material information of $P_c$ on $\hat{O}_c$ for regions with corresponding semantics. 
The synthetic exemplar \xy{$P_c$} has the corresponding ground truth semantic mask $P_s$ and part material assignment $\{s_P^i,  m_P^i\}_{i=1}^{n_P}$, where $s_P^i$ is a  part with a specific semantic label.
For the given projection $O_s$ with semantic parts $\{s_O^j\}_{j=1}^{n_O}$, since the paired \xy{colored} image \xy{$\hat{O}_c$} is translated from \xy{$P_c$} and the goal is to transfer the part material based on the semantic mapping between $P_s$ and $O_s$, we set the ground truth part material assignment for \xy{$\hat{O}_c$} to be $\{p_O^j, m_O^j\}_{j=1}^{n_O}$, where $m_O^j =m_P^i $ if there exists some $i$ such that $s_O^j = s_P^i$, and use only the set of parts with corresponding material assignment to fine-tune the material prediction network.
Otherwise, we set the ground truth as the material from the set $\{m_y^i\}$ with minimal perceptual distance to the predicted material.
Figure~\ref{fig:assign_part} illustrates this method. 
%which has ground truth segmentation and corresponding material assignment, we can use the 

%For example, for the purple part in $O_s$ without matched part in $P_c$, we cannot define the ground truth material for the corresponding region in the translated \xy{color} image  $\hat{O}_c$

%For ($\hat{O}_c, O_s$) pairs, 

\subsubsection{Fine-tuning with translated segmentation.} 
For ($\xy{P_c}, \hat{P}_s$) pairs, as the generated segmentation $\hat{P}_s$ is translated from $O_s$, when \xy{$P_c$} has parts that do not exist in $O_s$, the translated  $\hat{P}_s$ may become under-segmented, where some semantic parts could cover regions with multiple different materials and thus cannot be assigned with a single material label as ground truth. 
To overcome this problem, we oversegment the exemplar $P_c$ and use its intersection with the translated segmentation $\hat{P}_s$ as the part masks. 
More specifically, each semantic part $s_{\hat{P}}^{i}$ of $\hat{P}_s$ is subdivided into a set of smaller regions $\{s_{\hat{P}}^{\{i,t\}}\}_{t=1}^{n_{i}}$,  and each over-segment $s_{\hat{P}}$ is paired with the corresponding \xy{colored} region in \xy{$P_c$} to be the input to the material prediction network. The dominant material of the paired \xy{color} region is taken as the ground truth.
We find that, with the over-segmentation, the part materials become more consistent, and the training data obtained is more reliable for fine-tuning the material prediction network.
}

%\rh{
%[may need to add a figure to illustrate the process]
%
%}

\subsubsection{Fine-tuning with consistency loss.}
\change{
	To ensure that the material assignments predicted from the two image pairs obtained by the image translation are consistent during the fine-tuning, we add a consistency loss $L_c$.
	Note that for ($\xy{P_c}, \hat{P}_s$) pairs, each semantic part $s_{\hat{P}}^{i}$ may be subdivided into multiple regions $\{s_{\hat{P}}^{\{i,t\}}\}_{t=1}^{n_{i}}$ for material prediction, as explained above, and thus there may exist a one-to-many mapping between semantic parts from ($\xy{\hat{O}_c}, O_s$) and ($\xy{P_c}, \hat{P}_s$). 
	Our goal is to minimize the weighted average perceptual distance for the material predicted for the parts from the two different pairs but with the same semantics. More specifically, the consistency loss is defined as:
	\begin{equation}
		\label{eq:cons}
		 L_c  =  \sum_{i_k, j_k}\left(    \sum_{t=1}^{n_{j_k}}  \omega_t  L_{\text{dis}}(m_O^{i_k},  m_P^{\{j_k, t\}}) \right),
	\end{equation}
	where $\{i_k, j_k\}$ is the set of matched part indices between two pairs, $n_{j_k}$ is the number of sub-regions of semantic parts $s_{\hat{P}}^{j_k}$ of $\hat{P}_s$ after over-segmentation, $m_O^{i_k}$ and $m_P^{\{j_k, t\}}$ are the material prediction results for the corresponding parts of pair ($\xy{\hat{O}_c}, O_s$)  and pair ($\xy{P_c}, \hat{P}_s$), respectively, $L_{\text{dis}}$ is the perceptual distance loss defined in Eq.~\ref{eq:dist}, and $\omega_t$ is the normalized weight over part $s_{\hat{P}}^{j_k}$ based on the area of subdivided regions.
}

\begin{figure}[!t]
	\centering
	\includegraphics[width=\linewidth]{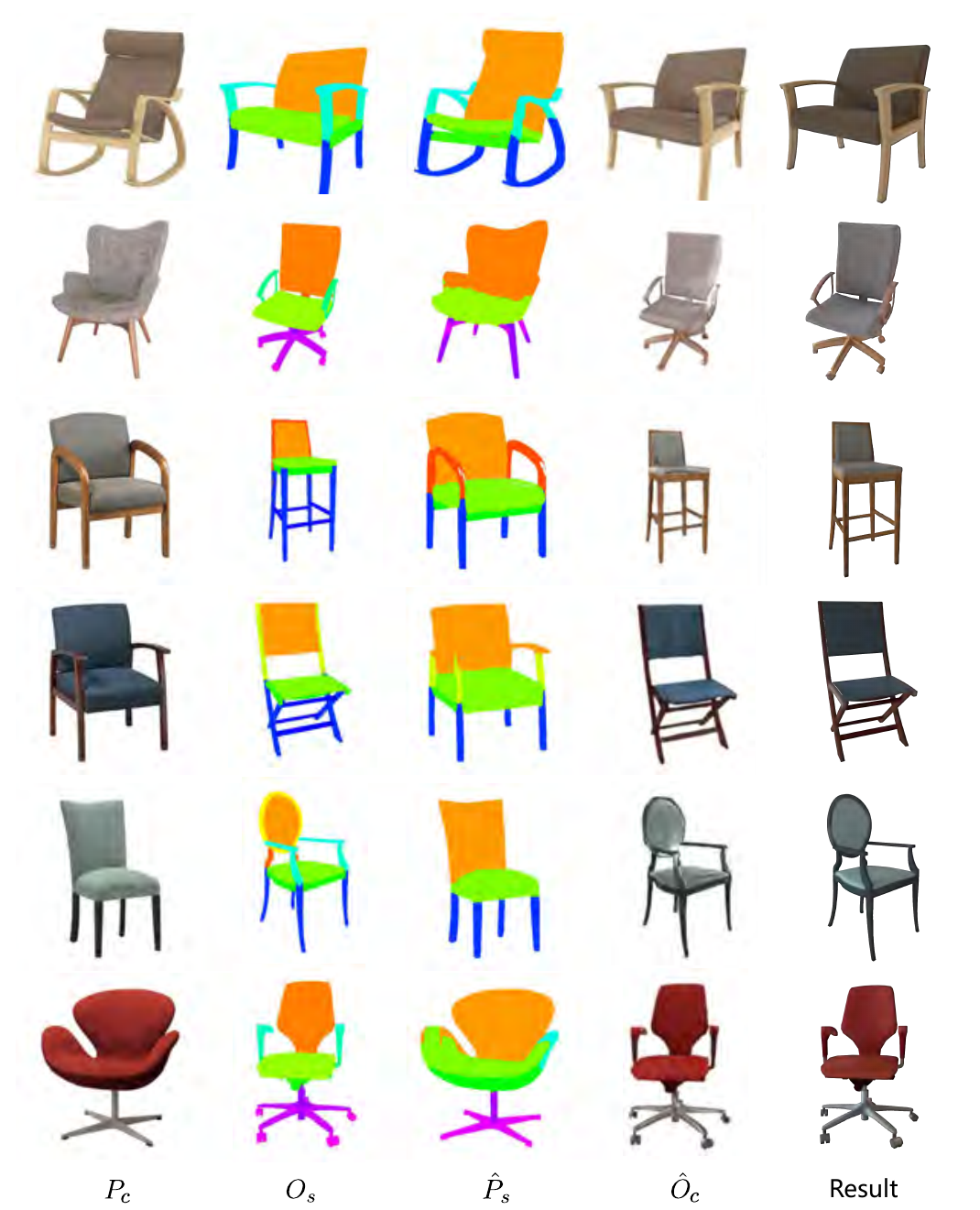}
        \caption{\revision{Sample of results showing the steps of our method: Input exemplar and segmentation (left), segmentation and \xy{color} translation (middle), and material transfer result (right). Rows 1-2: shapes with similar number of parts but with different structure; Rows 3-4: target shapes with fewer parts; Rows 5-6: target shapes with more parts.}}
	\label{fig:method_step}
\end{figure}

\begin{figure}[!t]
	\centering
	\includegraphics[width=\linewidth]{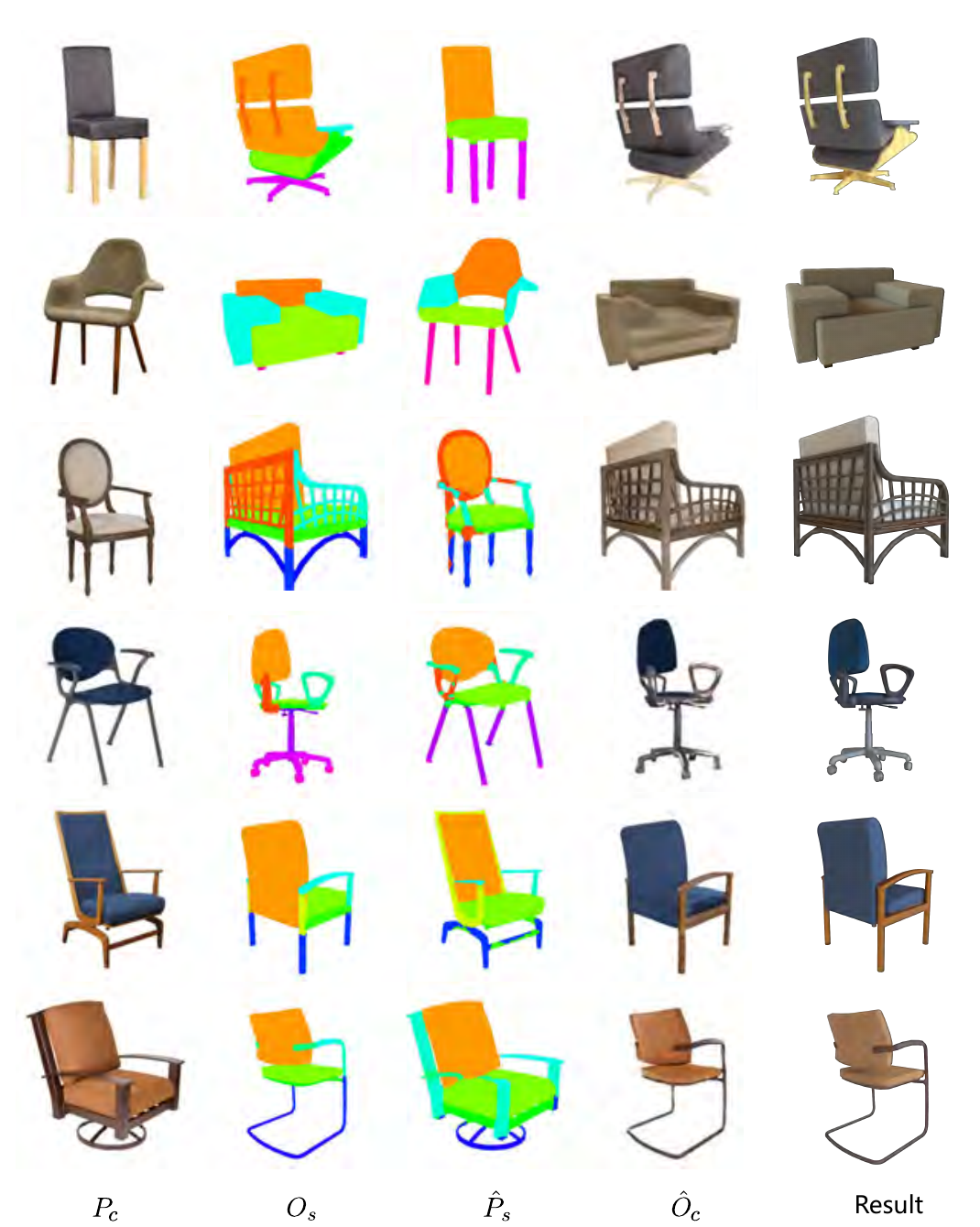}
	\caption{\revision{Our material transfer method is robust and generates high-quality results even in cases of incorrect camera pose estimation. The figure shows the input exemplar and segmentation (left two columns), segmentation and \xy{color} translation (middle two columns), and the material transfer result (right column).}}
	\label{fig:camera}
\end{figure}

\section{Results and Evaluation}
\change{In this section, we first present qualitative results and comparisons of our method. Then, we perform a set of experiments to evaluate the results and different components of our method in a quantitative manner, showing the importance of each component.
The set up for training our method can be found in the supplementary material.}

%\subsection{Training details}
% \rh{[May need to add some training details, like  parameter settings, computer info, the running time, training epochs, and model capacity.]}
%We first pre-train the image translation and material prediction networks on our synthetic dataset, and then fine-tune the full pipeline after combining the two networks together, as described in Section~\ref{sec:method}. 
%In this section, we present qualitative and quantitative evaluations of our method.

\subsection{Qualitative results}

%Figure~\ref{fig:biggallery} shows a sample of PhotoShapes generated with our method, to illustrate that our material transfer method can be used to generate collections with a variety of PhotoShapes. \ov{All these examples use real photos as input exemplars.} 
\change{\xy{Figure~\ref{fig:teaser}} shows a gallery of results generated with our method, to illustrate that our material transfer method can be used to generate collections with a variety of PhotoShapes. All these results were generated from \revision{in-the-wild photos with foreground extracted using automatic segmentation. Specifically, we used the Kaleido background removal plugin from Photoshop (version 2.0.6).}}

%\rh{[need to update the text accordingly.]}
\change{Figure~\ref{fig:method_step} shows a sample of results obtained with our method, where we show the input exemplar and segmentation on the left, translated segmentation and \xy{color} images in the middle, and final material transfer result on the right.
%the input of the method on the left, intermediate steps of the method in the middle, and two views of the result on the right. 
When inspecting these results, we see that the method is able to handle chairs with different structure and geometry. \revision{For example, the method successfully transfers the materials from a square to a round back (row 5), exchanges materials between legs with different topologies (rows 2 and 3), armrests with different topologies (row 2), and backs with different topologies (row 4). The presence of extra parts such as armrests or leg supports does not affect the transfer results (rows 3 and 6).} In addition, the method is also successful in assigning materials to chairs with arm rests even though the exemplars do not have this part (rows 5 and 6). In these cases, the method is able to infer the material from the semantic labels of the parts. Note that previous methods that assume sufficient similarity between the shape and exemplar would fail in most of these difficult cases.}

%On the right half of the gallery, we see examples where the 3D shapes and the objects in the exemplars have larger structural differences. Our method also produces PhotoShapes of good quality for these examples. For example, the method is able to match legs with very different topology and successfully transfer the materials (rows 1--5, right), performing similarly for backs of different topology (rows 2, 3, 9, right) and for backs with large geometric difference (rows 4 and 8, right). The 

\change{Figure~\ref{fig:camera} shows results obtained with incorrect camera pose estimation, where the most common error in the camera pose estimation step is a type of symmetric reflection of the correct pose. We find that our method still generates high-quality results despite incorrect camera pose estimation and imperfect translated segmentation $\hat{P}_s$. Since our goal is to estimate the material for each semantic part in $O_s$, as long as enough parts are visible in the selected pose, the image translation method is able to predict proper materials for the target parts.}
%is stable with the incorrect camera pose estimation and impeerfect translated segmetation $\hat{P}_s$ as our goal is the esitmate the material for each semantic part in $O_s$. ...}

\begin{figure}[!t]
	\centering
	\includegraphics[width=0.96\linewidth]{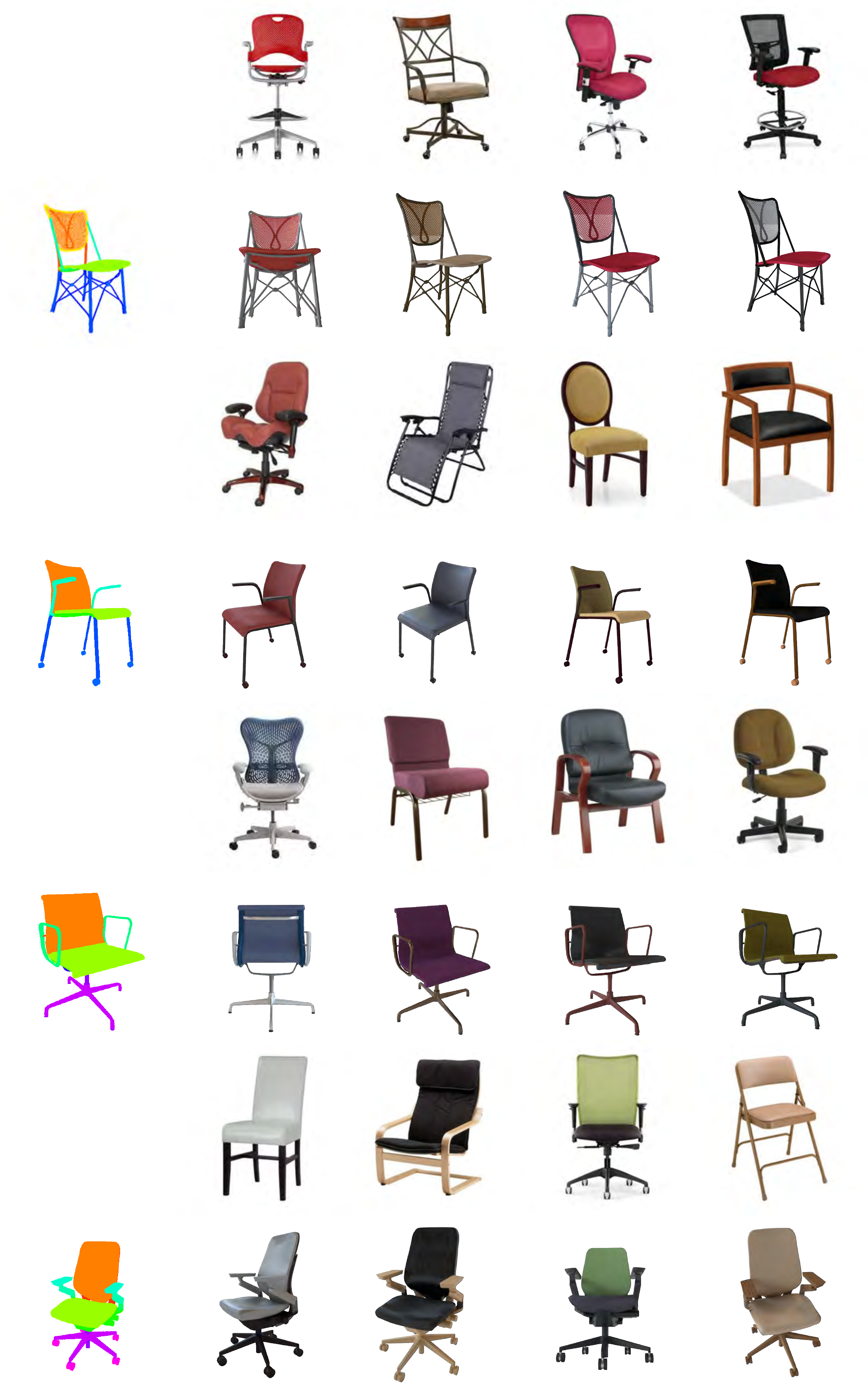}
        \caption{\change{Example results where the material of different exemplars (columns) is transferred to the same shape (rows). For better comparison, we rotate the 3D shape to the same view as the exemplar.}
         %\rh{[need to update.]}
     }
	\label{fig:diff_exemplar}
\end{figure}

\begin{figure}[!t]
	\centering
	\includegraphics[width=0.96\linewidth]{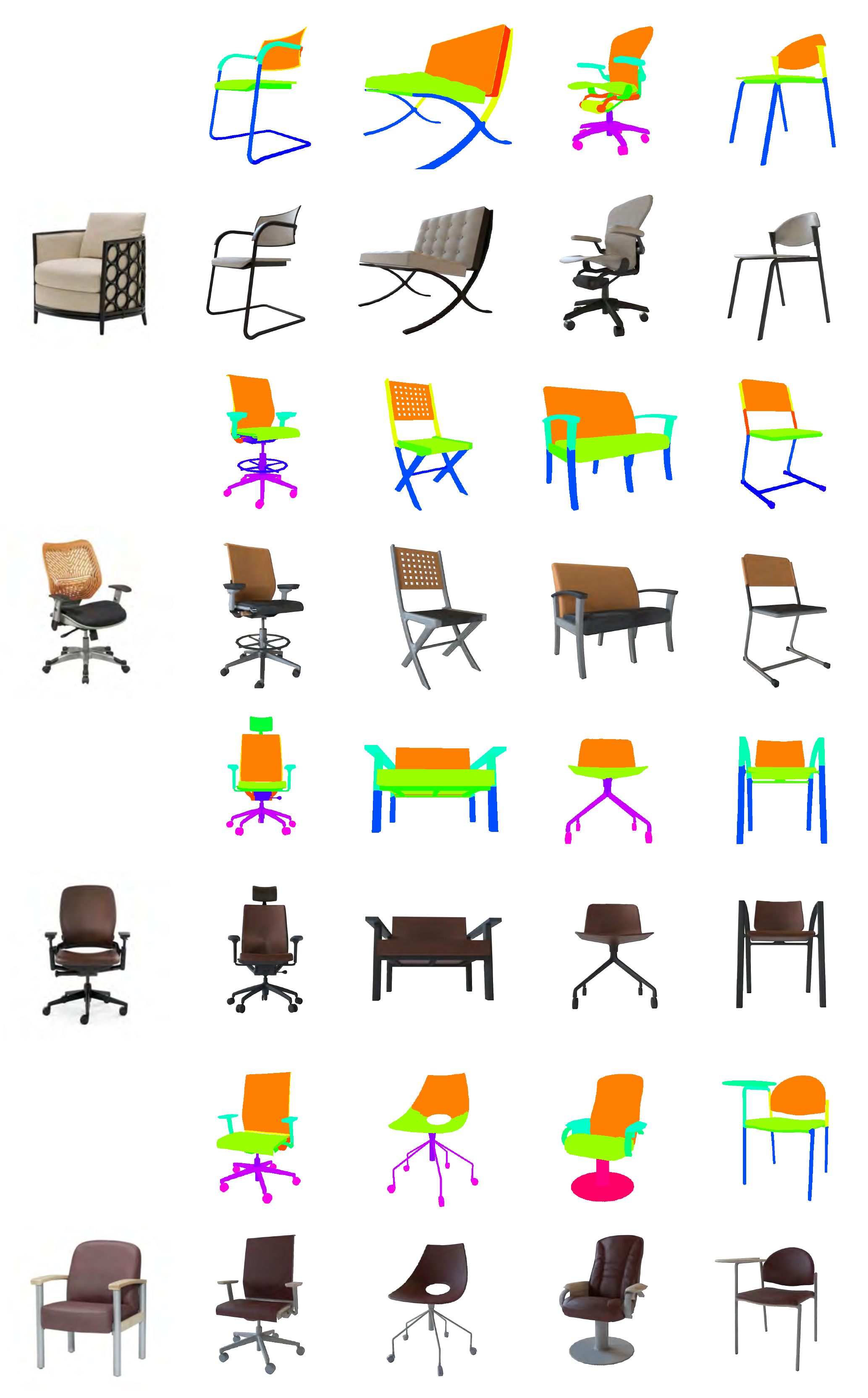}
	\caption{\change{Example results where the material of the same exemplar (rows) is transferred to different shapes (columns). For better comparison, we rotate the 3D shapes to the same view as the exemplar.}
	% \rh{[need to update.]}
 }
	\label{fig:diff_shape}
\end{figure}

%\rh{ [Figure~\ref{fig:diff_exemplar} and Figure~\ref{fig:diff_shape}  have been updated.]}
Moreover, Figure~\ref{fig:diff_exemplar} shows examples of results where we fix a 3D shape and transfer the materials from different exemplars to the same shape. We see that the method can transfer materials from different \xy{categories} to the same shape, such as woods, metals, and leathers.
 Figure~\ref{fig:diff_shape} shows the complementary scenario where we fix an exemplar and transfer it to different 3D shapes. With these two comparisons, we see again that the structure and geometry of the exemplar does not need to perfectly match those of the 3D shape, where both the shape and exemplar can have additional parts that do not appear in the other object. 
 The method is robust to these shape differences and thus can be applied to a great variety of objects. In addition, we see from these two comparisons that the user can choose a specific exemplar and 3D shape for the material assignment. Previous methods are able to transfer materials only when the exemplar and shape are sufficiently similar, implying that the user has less control over the input.

%\ov{[Add results with real photos if possible]}

\subsection{Qualitative evaluation}
\change{As there is no ground truth for the material transfer results due to the difficulty involved in its creation, we conduct a user study to evaluate the quality of the results generated by our method comparing to the results of baseline methods.}

\change{
\paragraph{Baseline methods.}
We compare our method to two different baselines that solve the same problem as our method with different strategies. 

The first baseline is the \textit{PhotoShape} method of Park et al.~\shortcite{park2018photoshape}, which is the previous work most related to our method. We run PhotoShape using the authors' code for shape-image alignment, but our network for material prediction, since we show in Section~\ref{sec:quantitative} that the material prediction accuracy of our method is higher. 

The second baseline, denoted as \textit{PhotoShape+}, is a simplified version of our method, which incorporates a segmentation network with the shape-image alignment proposed in~\textit{PhotoShape}~\cite{park2018photoshape}.
More specifically, instead of using a translation network, we train an image segmentation network to first get the semantic segmentation of the photo exemplar. Then, we provide the photo exemplar and each of its predicted semantic parts as input to our material prediction network to obtain the part material labels. 
We use DeepLabv3+~\cite{chen2018encoder} as the segmentation network. 
Note that for this baseline, we can only assign materials to shape parts that have corresponding parts in the photo exemplar. 
%\revision{There might be ways to assign material for unmatched part, for example, renormalize the predicted distribution using the semantic labels that are in the 3D shape or assign the semantic labels based on similarities for missing parts, but we found that such renormalization don't lead to meaningful results for most of the cases and it is difficult to justify the handcrafted equivalence rules for different parts.}
\revision{It may be possible to use heuristics for assigning materials to unmatched parts, for example, by renormalizing the predicted distribution using the semantic labels that exist in the 3D shape, or assigning the semantic labels based on similarities to missing parts. However, we found that renormalization does not lead to meaningful results for most of the cases, and thus it is difficult to justify using handcrafted equivalence rules for different parts.}
Thus, for any unmatched part, we deform $O_s$ to align it with the photo exemplar \xy{$P_c$} using the shape-image alignment method proposed by Park et al.~\shortcite{park2018photoshape}, and then assign the material of the corresponding region in the photo exemplar to the part to get a complete material assignment for the given shape. 
}
%Then, we provide the photo exemplar and its predicted segmentation as input to the material prediction network to obtain the part material assignment. 
%\rh{Note that we can assign a material to a shape part if there exists a corresponding part in the photo exemplar, thus this baseline cannot generate complete final result, and we can only evaluate the material prediction accuracy for this method.}
%We assign a material to a shape part if there exists a corresponding part in the photo exemplar. Otherwise, we use different heuristics to assign appropriate materials to the parts without matches. The different heuristics provide two different baselines that we compare: \textit{Baseline 1}: rule-based. The part labels form a pre-defined semantic hierarchy, and thus we can use such hierarchical information to find the most related part in the shape that has a material assignment. Then, we copy the material of the related part to the part without a match to the exemplar. This process is similar to the process described in Section~\ref{sec:datasets} for creating our training data. \textit{Baseline 2}: deformation-based. We deform the 2D projection of the 3D shape to align it with the photo exemplar using the shape-image alignment method proposed by Park et al.~\shortcite{park2018photoshape}. Then, after deformation, we simply assign the material of the corresponding region in the photo exemplar to the part. 

%\rh{[Do you think it's better to organize the details of user studies one by one, instead of discuss the setting together first and then results?]}

\change{
\paragraph{User study.}
%Our user study is composed of two sets of questions. 
    Before starting the user study, we explain the goal of material transfer to the users. The specific wording used in the instructions and in the questions is provided in the supplementary material. Then, we ask the users multiple questions where we show the result of one of the baselines and our method, in random order, and ask the user to select which result they think is better. The user can select either one of the two results or a ``not sure'' option.}

%The second set of questions is similar to the setting described by Park et al.~\shortcite{park2018photoshape}. We randomly show a result of our method, one of the baselines, or a real exemplar photo, and ask the user to select whether the image is a ``real photograph'' or ``generated by a computer''. The user can chose one of the two options or a ``not sure'' option. 

\change{To conduct the user study, we collected 300 sets of results by sampling exemplar images and 3D shapes from our dataset, and provided these as input to the different methods to obtain three different transfer results. After that, we compare our method to each baseline separately, which results in 600 questions. 
We asked 12 participants to do the user study, all of whom are graduate students in computer science. We collected 150 answers from each participant and thus 1,800 answers in total, with each of the 600 questions having answers from three different participants.
\revision{For each question, we take the answer selected by the majority of the participants as the final answer, and consider the answer to be ``not sure'' if there is no agreement among the answers when we get three different answers.}}
%each of the options is selected once. 

%For the second user study, 

%\input{figures/tab_comp_baseline}
\begin{figure}[!t]
	\centering
	\includegraphics[width=\linewidth]{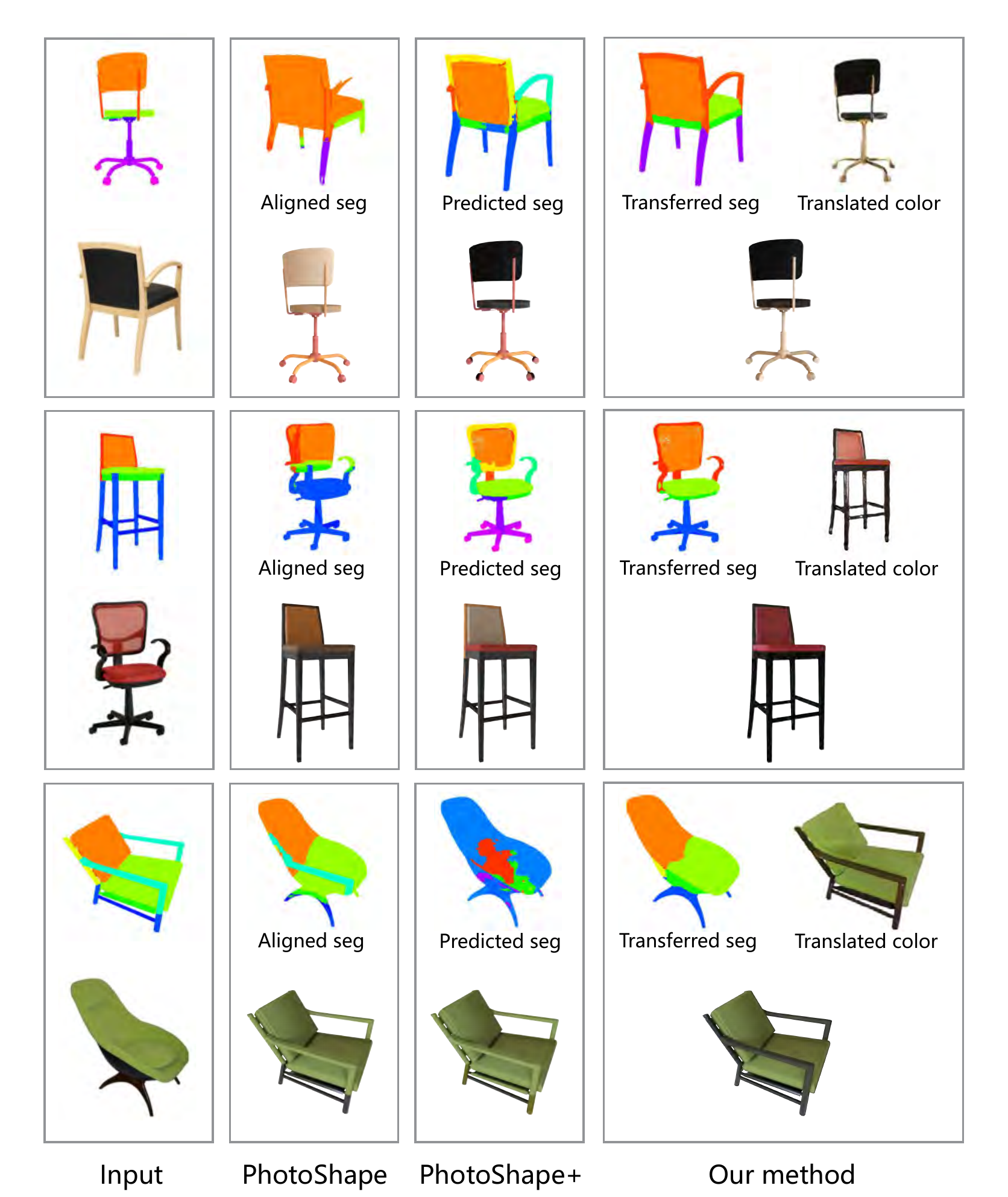}
	\caption{
            \change{Comparison of our method to two baselines. Note
        how the intermediate results of our method build a better correspondence between the input segmentation and exemplar, and how the final results of our method better resemble the exemplar. }
        %\rh{[may need to add more explanation on the intermediate results we shown here ]} 
    }
	\label{fig:comp_baseline}
\end{figure}

\change{
\paragraph{Results.}
%\ov{I assume the following has to change after the user study results.}
When comparing to PhotoShape, the vote percentages for the options ``PhotoShape/ours/not sure'' are $19.3\%/70.7\%/10\%$, and when comparing to PhotoShape+, the vote percentages for the options ``PhotoShape+/ours/not sure'' are $21.3\%/62.7\%/16\%$. We see that our method was selected much more frequently than other methods as having the best result, and we can also conclude that there must be a noticeable improvement in our results compared to other methods, given that the number of ``not sure'' votes is small.}
%\revision{To get a quantitative evaluation of how the final results obtained by different methods resemble the input photo exemplars, we also compute the Fréchet Inception Distance (FID) ~\cite{heusel2017gans} between the rendered images from the predicted camera views and the input photo exemplars as in ~\cite{oechsle2019texture}. The FID for ``PhotoShape/PhotoShape+/PhotoShape++'' are $58.66/58.16/57.33$, where the conclusion is consistent with human judgement.}
\revision{To evaluate in a quantitative manner how well the final results obtained by different methods resemble the input photo exemplars, we also compute the Fréchet Inception Distance (FID)~\cite{heusel2017gans} between the images rendered from the predicted camera views and the input photo exemplars, as in the work of Oechsle et al.~\shortcite{oechsle2019texture}. The FIDs for ``PhotoShape/PhotoShape+/ours'' are $58.66/58.16/57.33$, which are consistent with the human judgment.}

%\xy{[More analysis of the results.]}

\change{In Figure~\ref{fig:comp_baseline}, we present a visual comparison of our method to the two baselines on a few examples to provide more insight on the user study results.} 
\change{For each example, we show the intermediate results of all the methods for more detailed comparison. 
More specifically, we show 	the \emph{aligned segmentation} for PhotoShape, additional \emph{predicted segmentation} for PhotoShape+, and both \emph{translated segmentation} and \emph{\xy{color}} for our method.% (PhotoShape++). 

For example, when analyzing the results shown in the first row of the figure, we see that since the structures of the input 3D shape and the shape in the exemplar photo are quite different, especially with many thin parts, the segmentation obtained by the shape-image alignment method in PhotoShape is noisy, which results in image regions consisting of several different materials that correspond to a single semantic part. Thus, this segmentation leads to unreliable material prediction for parts like the back and seat of the chair.
Since PhotoShape+ relies on a more accurate semantic segmentation predicted by the segmentation network, PhotoShape+ builds a correct, clean correspondence between the semantic parts and the corresponding regions in the exemplar photo, and transfers the correct material to the seat and back of the chair. 
However, since the legs of the swivel chair have thin parts that do not exist in the exemplar shape, the material assignment is unsatisfactory, since the corresponding materials are assigned based on the alignment results obtained in PhotoShape, which are combined together to provide the final result for PhotoShape+.
Comparing to these two baselines, our method translates the segmentation from the input shape to the exemplar smoothly and in turn provides much better material assignments in the final results. Similar results can be found for the example shown in the second row.

For the result shown in the last row, when the exemplar has a smaller number of semantic parts than the 3D shape, the shape-image alignment method in PhotoShape  assigns materials predicted for parts with different semantics, which leads to unrealistic results, e.g., the green armrest. 
Moreover, as the chair in this exemplar photo has an irregular shape, the segmentation network used in PhotoShape+ outputs highly incorrect results which lead to incorrect material assignments for parts like legs. 
In contrast, our method benefits from the image translation to obtain a more accurate segmentation, and synthesizes a realistic \xy{color} image for material prediction for unmatched parts, i.e., the armrest, which together leads to the result that better resembles the exemplar. 

Overall, our results better conform to the materials seen in the input exemplar, both in \xy{category} and albedo. We see significant improvement especially when the matching parts have large geometric differences, or with missing parts.}
%The comparison of the baselines is shown in Table~\ref{tab:comp_baseline}, while a visual comparison of example results is shown in Figure~\ref{fig:comp_baseline}.
% Since the results of Baseline 2 are similar to those of Baseline 1, but with a slight improvement over Baseline 1, we only show the visual results of Baseline 2 in the figure. 
%In Table~\ref{tab:comp_baseline}, we see that our method obtains the best results according to the three measures, with an improvement of around 9 points for material accuracy when compared to the next best performing method (PhotoShape). The improvement of our method over \textit{PhotoShape+} is around 13 points, showing the importance of the image translation network. Note also the higher accuracy of \textit{PhotoShape} over \textit{PhotoShape+} , showing the difference between using a segmentation network or alignment method.
%The other two evaluation measures show similar trends.

%For the second user study, ...

\subsection{Quantitative evaluation}
\label{sec:quantitative}
\change{In this section, we conduct a quantitative evaluation of the material prediction network with synthetic image pairs, which provide ground truth material and \xy{category} labels that we can use for computing accuracy measures.}
%We perform two ablation studies to demonstrate the importance of the different components of our method.}

We evaluate the quality of the material labeling compared to the ground truth with three measures: 1 and 2. Material and \xy{category} accuracies (Mat-acc, Cat-acc, respectively), computed as the classification accuracy (number of correctly assigned labels / total number of parts), where the goal is to maximize the accuracy; 3. Material perceptual distance (Mat-dis), computed with the L2-lab distance~\cite{sun2017attribute}, where the goal is to minimize the distance between the ground truth and assigned materials.

\change{To prepare the synthetic (\xy{color}, segmentation) pairs, we first create (\xy{color}, segmentation) tuples by rendering the 3D shapes in our dataset from different views and extracting the corresponding semantic projections for each view based on the part segmentation of the shapes. Then, for each rendered image, we use the camera pose prediction network to predict the camera parameters of the image. To generate a second (\xy{color}, segmentation) tuple corresponding to the rendered image, we arbitrarily select another shape from the dataset and generate its rendering and semantic projection from the predicted view. This procedure provides us with translated (\xy{color}, segmentation) image pairs. We divided the image pairs into training and test sets. The details about the ground truth material assignment, synthetic image generation, and exact data division for training and testing can be found in the supplementary material.}
%The exact data division can be found in the supplementary material.

\change{Note that the material prediction network is first trained using \emph{synthetic} (\xy{color}, segmentation) image pairs, which are obtained by directly rendering the 3D shapes in our dataset with assigned materials from different views, and then fine-tuned with \emph{translated} (\xy{color}, segmentation) image pairs obtained with our image translation method by taking an exemplar \xy{color} image and a 3D shape as input. 
We have ground truth material labels for all the \emph{synthetic} image pairs, but only partial ground truth for \emph{translated} images pairs, which is derived from  the semantic correspondence between the input exemplar and 3D shape as discussed in Section~\ref{sec:mattransf}.
Thus, we perform two ablation studies to evaluate these two fine-tuning steps separately to demonstrate the importance of considering the perceptual similarity for material prediction and the importance of ensuring the consistency between the prediction results of two translated image pairs for material transfer. 
%those demonstrate the importance of the different components of our method.
}

%All the experiments described below use synthetic images as exemplars, 

%\rh{[need to update the following metric]}
%After computing the material assignments with our method and the three baselines, 

% So, the main difference is the segmentation network versus alignment
% method, so no need to talk about the alignment of missing parts.
%
% although both baselines used the same alignment method, PhotoShape provides a higher accuracy since all of the material assignments are based on the alignment, while Baseline 2 uses the alignment only for unmatched parts} \rh{and the low accuracy is mainly caused by the generally over-segmentation where \xy{color} with small part mask tends to get incorrect material type.}

%\ov{due to ...[do we really use the same image deformation/alignment method, or just similar but not the same? We can also remove this comment, but I thought it is better to comment it as it is quite noticeable.]} \rh{[For baseline 2, most of the part materials are determined by the semantic label and only the ones without matching are found based on the alignmented, while PhotoShape entirely based on the alignment result, that's why the perfermance of the baseline 2 is better than PhotoShape and the visual result is also better.]}

%We are also trying to add a user study to visually compare the results obtained by our method and baselines. \rh{[May not have time to do this, only show some visual examples instead.]}

\begin{table}[!t]%
    \caption{\change{Comparison of the material prediction network trained with different losses (Settings 1--3 in the text). Note the better performance of the full loss 	$ L_{\text{class}} +  L_{\text{metric}} $, where $L_{\text{class}} = \xy{L_{\text{cat}}} +  L_{\text{mat}}  +  L_{\text{dist}}$ is the classification loss and $L_{\text{metric}} $ is the metric learning loss.}} 
	\label{tab:comp_prediction}
	\begin{minipage}{\columnwidth}
		\begin{center}
			\begin{tabular}{l|l|l|l}
				\hline\hline
				\textbf{Loss}& \textbf{Mat-acc ($\uparrow$)}&  \textbf{Sub-acc($\uparrow$)}  &   \textbf{Mat-dis ($\downarrow$)}  \\ \hline \hline 
				$\xy{L_{\text{cat}}} +  L_{\text{mat}} $  & 57.53\% & 81.89\% &  7.69 \\  \hline 
				$\xy{L_{\text{cat}}} +  L_{\text{mat}}  +  L_{\text{dist}}$ & 61.56\%  & 84.30\% & 6.38  \\  \hline 
				%$L_{\text{class}} $ & 61.56\%  & 84.30\% & 6.38  \\  \hline 
				$ L_{\text{class}} +  L_{\text{metric}} $ & \textbf{71.62\%} & \textbf{86.49\%} & \textbf{4.66} \\ \hline \hline 
			\end{tabular}
		\end{center}
	\end{minipage}
\end{table}%

\begin{figure}[!t]
	\centering
	\includegraphics[width=\linewidth]{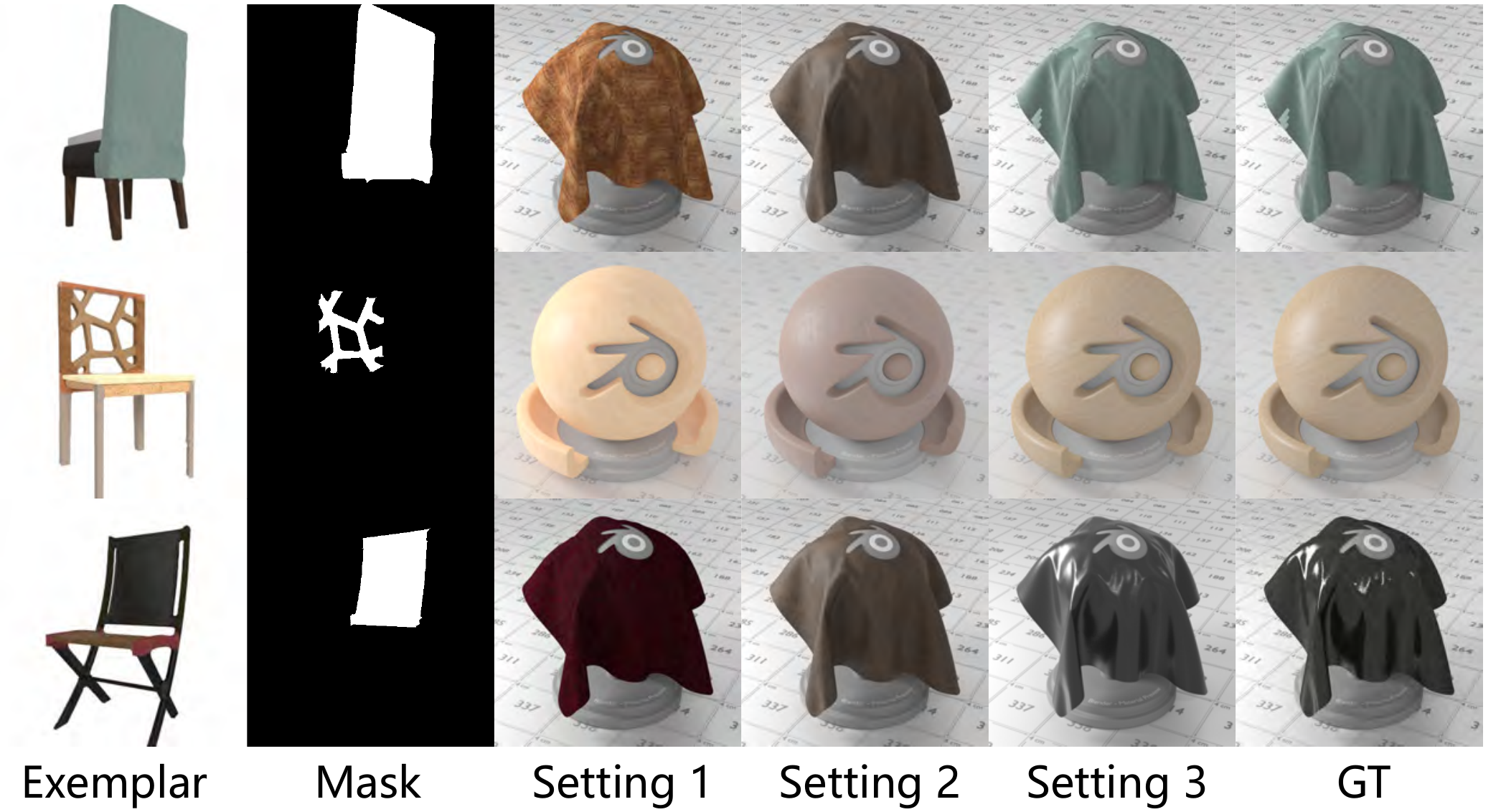}
	\caption{
        Examples comparing the material prediction network learned with different settings of the loss function. We use Setting 3.}
	\label{fig:comp_pred}
\end{figure}

%\paragraph{Ablation studies.} 
\change{
%\textit{Evaluation with synthetic data having ground truth.} 
\textit{Importance of considering perceptual similarity. } 
We first use synthesized (\xy{color}, segmentation) image pairs to compare the material prediction network learned with and without the perceptual similarity, and also evaluate the learning of the perceptual similarity with different forms of the loss function. We evaluate the following settings: \\
%\begin{itemize}
$\bullet$ \textit{Setting 1}: the prediction network trained without the perceptual distance loss (the same setting as Park et al.~\shortcite{park2018photoshape}). This corresponds to training the network with the loss $\xy{L_{\text{cat}}} + L_{\text{mat}}$; \\
$\bullet$ \textit{Setting 2}: directly adding the perceptual distance loss to the classification loss, i.e., using $\xy{L_{\text{cat}}} + L_{\text{mat}} + L_{\text{dis}}$; \\
$\bullet$ \textit{Setting 3}: using the perceptual distance measure in a softer way to learn a perception-aware feature space first, i.e., using the distance matrix to sample triplets to perform metric learning, and then fine-tuning the network to perform the classification with the same loss as \textit{Setting 2}. Thus, Setting 3 uses all the terms of the loss function, i.e., $L_{\text{metric}} + L_{\text{class}} $, and corresponds to the setting used in our method.
%\end{itemize}

%combining the perceptual distance with the classification task, i.e., using $ L_{\text{sub}} + L_{\text{mat}} +  L_{\text{tri}} + L_{\text{sim}}$.
% \textit{Setting 4}: using all terms of the loss function, i.e., $ L_{\text{sub}} + L_{\text{mat}} + L_{\text{dis}} +  L_{\text{tri}} + L_{\text{sim}} $, which corresponds to the network used in our method. \rh{[still training...]}

The comparison is shown in Table~\ref{tab:comp_prediction}, while a visual comparison is shown in Figure~\ref{fig:comp_pred}. With this study, we confirm that the use of the perceptual loss and the full loss function provide the best material assignments, where the accuracy increases with the addition of each term of the loss function. In Figure~\ref{fig:comp_pred}, we see the same gradual improvement with the addition of more terms of the loss.

%\textit{Evaluation with translated pairs.} 
\textit{Importance of fine-tuning with consistency loss. } 
For the two (\xy{color}, segmentation) image pairs generated by the image translation step, i.e., ($\xy{P_c}, \hat{P}_s$) and ($\xy{\hat{O}_c}, O_s$), we show the necessity of the fine-tuning process described in Section~\ref{sec:mattransf}, which fine-tunes two separate networks and also uses a consistency loss. 
We compare the material prediction accuracy before and after fine-tuning for both pairs in Table~\ref{tab:comp_path}. \change{Figure~\ref{fig:comp_path} shows a visual comparison of results.}
%For ($P_c, \hat{P}_s$) pairs, the fine-tuning mainly constrains the prediction results of ($\hat{O}_c, O_s$) pairs. Thus, we only compare the performance of the corresponding predictor $\text{Pred}_P$ before and after the common fine-tuning process using  ($P_c, \hat{P}_s$) pairs, denoted as $\text{Pred}_P$ and $\text{Pred}_P^{*}$, respectively.
%For ($\hat{O}_c, O_s$) pairs, since the fine-tuning can refine the results based on ($\hat{O}_c, O_s$) pairs or be constrained by the ($P_c, \hat{P}_s$) pairs with the use of the consistency loss, we compare four different fine-tuning options, denoting $\text{Pred}_O$ the predictor without fine-tuning, $\text{Pred}_O^{*}$ the predictor fine-tuned with ($\hat{O}_c, O_s$) pairs, and $\text{Pred}_O^{*} + \text{Pred}_P$ and $\text{Pred}_O^{*} + \text{Pred}_P^{*}$ the predictors fine-tuned with the consistency loss defined on the output from $\text{Pred}_P$ and $\text{Pred}_P^{*}$, respectively.

\begin{table}[!t]%
	\caption{\change{Comparison of the prediction accuracy for the translated image pairs before and after fine-tuning. Note the better performance with fine-tuning for both ($\xy{P_c}, \hat{P}_s$) and ($\xy{\hat{O}_c}, O_s$).}
			%different settings for using the two pairs of images output by the translation network to fine-tune the material predictor. 
		%	In our setting, we use two predictors that are first fine-tuned on two types of image pairs separately  and then combine them with the consistency loss to get the best results for ($\hat{O}_t, O_s$) pairs that we use to get the final results, as shown in the last row. \ov{[Should we change to Mat-a, Sub-a, and Mat-d so that it fits in the space?]}}
		%We use Pair 1 + 2 + $L_c$ in our method. %\rh{[Currently, we just succesfully fine-tune the translation network with consistency loss. To be able to fine-tune the prediction network with the consistency loss, we need to make significant change to the code and may run out of time. I will put the number we obtained by only fine-tuning the translation network with consistency loss for now, and the ony problem is that the substance accurarcy is a bit lower than the case without consistency loss.]}
	} 
	\label{tab:comp_path}
	\begin{minipage}{\columnwidth}
		\begin{center}
			\begin{tabular}{l|c|l|l|l}
				\hline\hline \textbf{Data} 
				& \textbf{Fine-tune} & \textbf{Mat-acc ($\uparrow$)}&  \textbf{Sub-acc($\uparrow$)} &   \textbf{Mat-dis ($\downarrow$)} \\ \hline \hline 
				
				\multirow{2}{*}{($\xy{P_c}, \hat{P}_s$)} 
				& N  & 64.87\% &  85.26\%  & 8.31 \\ \cline{2-5} 
				& Y  & \textbf{79.17\%} & \textbf{90.41\%} & \textbf{5.24}   \\ \hline \hline

				\multirow{3}{*}{($\xy{\hat{O}_c}, O_s$)} 
				& N & 10.26\%&  43.83\%  &  16.88 \\ \cline{2-5} 
				& Y  & 24.18\%  &  59.71\%  & 12.10  \\\cline{2-5} 
				& Y + $L_c $& \textbf{26.95\%} & \textbf{62.32\%} & \textbf{11.17}\\ \hline \hline
				
			\end{tabular}
		\end{center}
	\end{minipage}
\end{table}%

\begin{figure}[!t]
	\centering
	\includegraphics[width=\linewidth]{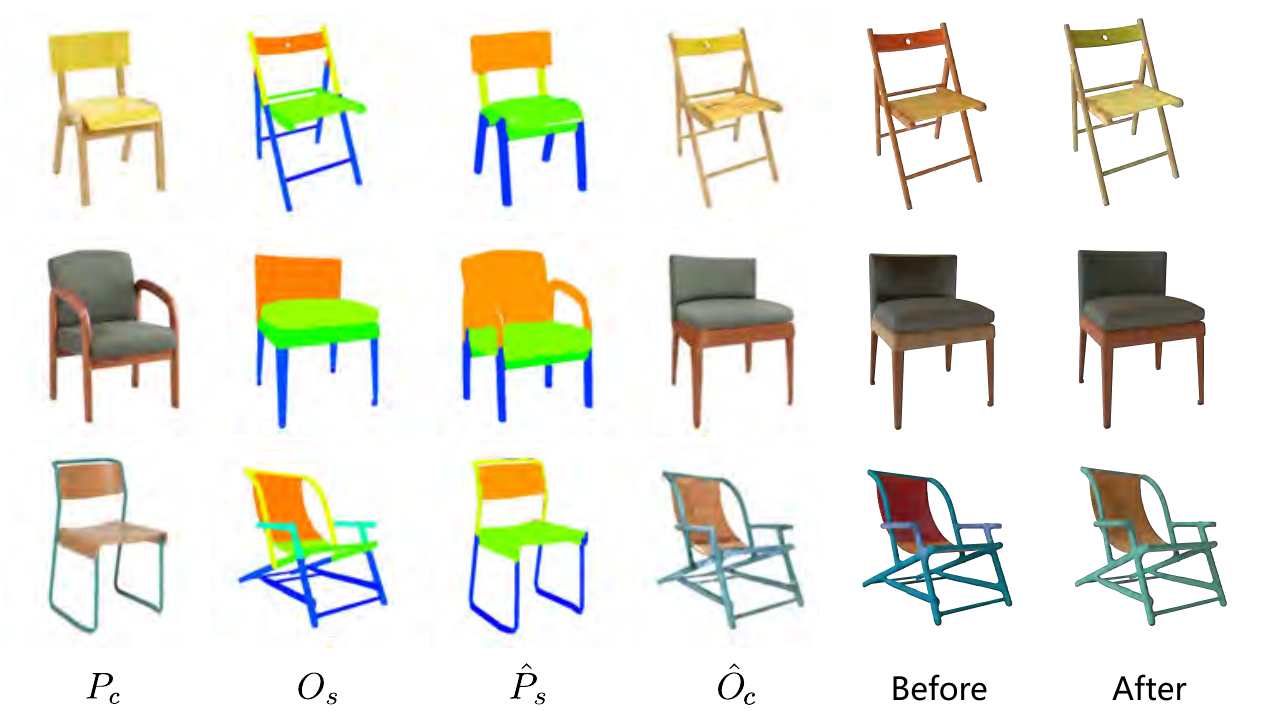}
	\caption{
		\change{Comparison of results before and after fine-tuning. Note how the results better resemble the exemplars $P_t$ when fine-tuning is used.}}
	\label{fig:comp_path}
\end{figure}

%The comparison is reported in Table~\ref{tab:comp_path}. 
When comparing the results from two different image pairs, we see that the prediction accuracy of the ($\xy{\hat{O}_c}, O_s$) pairs is much lower than that of ($\xy{P_c}, \hat{P}_s$) pairs. The reason is that the material prediction network takes a \xy{color} image and a mask as input, and the prediction results are highly determined by the quality of the \xy{color} image, while the material prediction network is trained on rendered data, which is quite different from the translated \xy{colored $\hat{O}_c$}.  
However, note that for parts that can be found in $\hat{P}_s$, we use the prediction result for the ($\xy{P_c}, \hat{P}_s$) pair, and the prediction results for ($\xy{\hat{O}_c}, O_s$) are only used for unmatched semantic parts.
We can see that %fine-tuning only with their own corresponding image pairs leads to the less accurate results.
by adding the consistency loss $L_c$ to ensure consistent prediction results for the corresponding two image pairs, the results are improved. 
Our current setting shown in the last row obtains the best results on all the three metrics.

}

\begin{figure}[!t]
	\centering
	\includegraphics[width=0.98\linewidth]{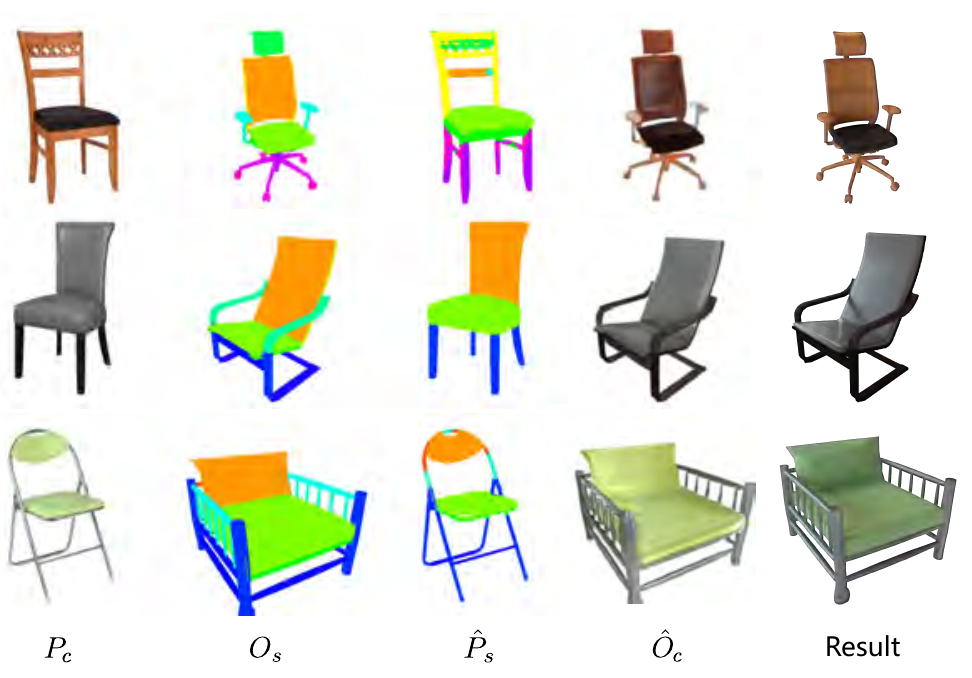}
        \caption{Representative failure cases of our method: \change{semantic mismatch between exemplar and shape (Row 1), incorrect material inference (Row 2), and insufficient variety of the material dataset (Row 3).}}
        %\ov{[In Row 1, the correspondence is indeed incorrect, but the result seems to make sense as the armrest would be similar to the leg? Row 2 does not look too bad... and is Row 3 about the metallic material?]} \rh{[Row 3 the material substance type is the same but we don't material with such color in the given exemplar, the green looks a bit different.]}}
	\label{fig:failure}
\end{figure}

\subsection{Failure cases}
%
%\rh{[May move to the end in the discussion of limitation and failure cases?]}
%\ov{[I think we can simply report the failures here, and then refer to
%        them in the conclusions for more discussion?]}
%    \rh{[Sounds good.]}

%\rh{[need to update the text accordingly.]}
\change{Figure~\ref{fig:failure} shows example results that represent the main failure modes of our method. We identify three main failures cases: 1)~The image translation and material assignment are correct. However, the result does not make sense due to a semantic mismatch between the exemplar and the shape. For example, in the first row of the figure, a wooden material is assigned to the back of a swivel chair. Such type of \xy{category} assignment is rarely found in real-world designs. 
	% 1) The image translation provides an incorrect correspondence between the exemplar and shape due to the large differences between the shapes, resulting in an incorrect material transfer. For example, in the first row of the figure, the armrests of the result have the material corresponding to different parts of the exemplar. 
%2) A shape part does not have a corresponding part in the exemplar and thus the incorrect material is assigned. For example, in the second row of the figure, the armrests are assigned the material of the exemplar's legs. 
2)~The material prediction network predicts a visually similar but incorrect material \xy{category}, such as the seat of the chair shown in the second row, where a metal material is assigned instead of a fabric material.}
\change{Note that the material predicted for the back is more accurate even though the materials presented in the exemplar photo are the same for the seat and the back.   
This incorrect material prediction is caused by different reflectance effects on different parts. Inconsistent prediction results for related parts are due to the independent per-part material prediction of our method.}
\change{3)~The material assignment does not match perfectly with the exemplar, due to insufficient variety of materials in the dataset. In the third row of the figure, the green materials have a slightly different hue.}
%3) The material assignment does not match perfectly with the exemplar, due to insufficient variety of materials in the dataset. In the third row of the figure, the metallic materials have a slightly different luminosity.
%The image translation and material assignment are correct. However, the result does not make sense due to a semantic mismatch between the exemplar and the shape. For example, in the first row of Figure~\ref{fig:failure}, a wooden material is assigned to the back of a swivel chair. Such type of substance assignment is rarely found in real-world designs.  }

\begin{figure}[!t]
	\centering
	\includegraphics[width=\linewidth]{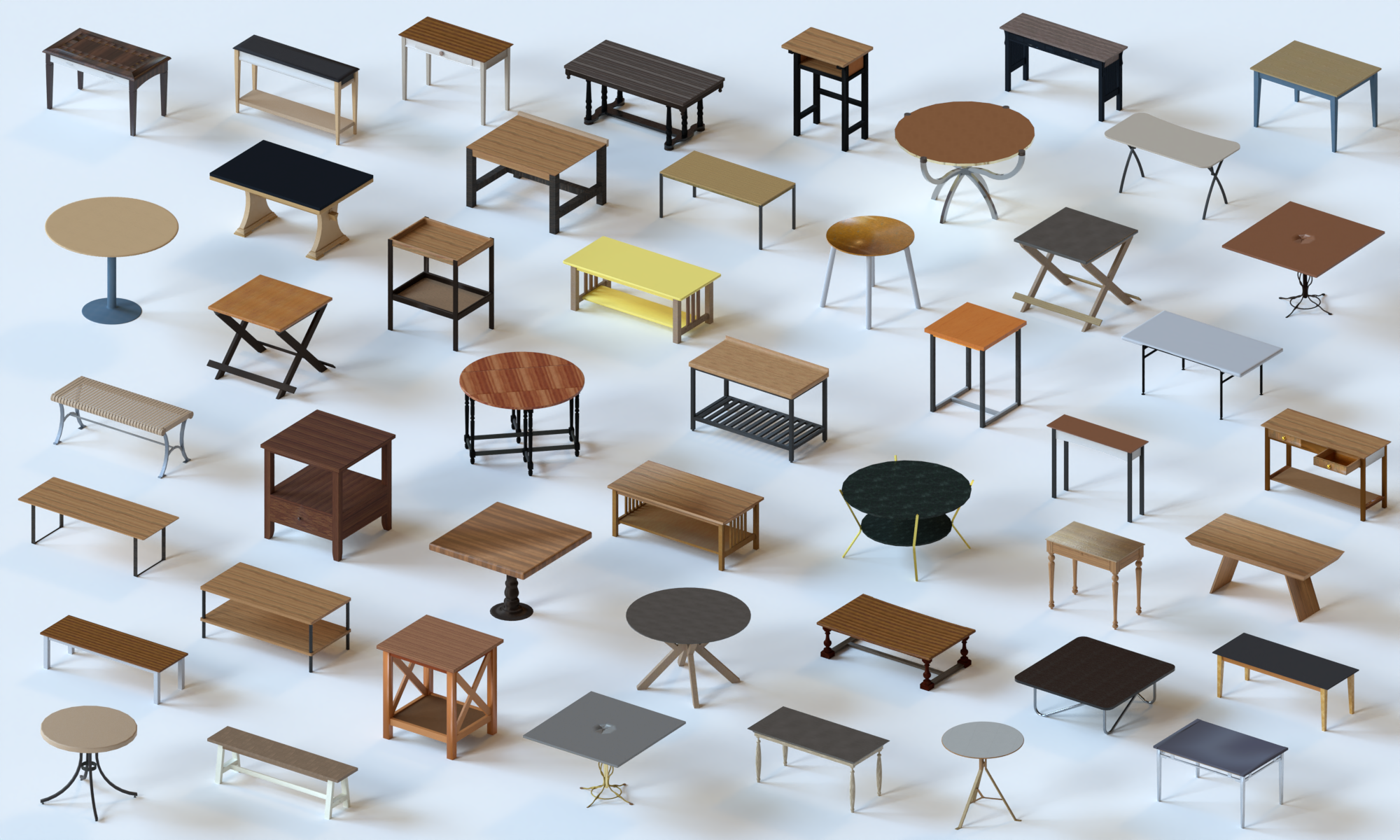}
	\includegraphics[width=\linewidth]{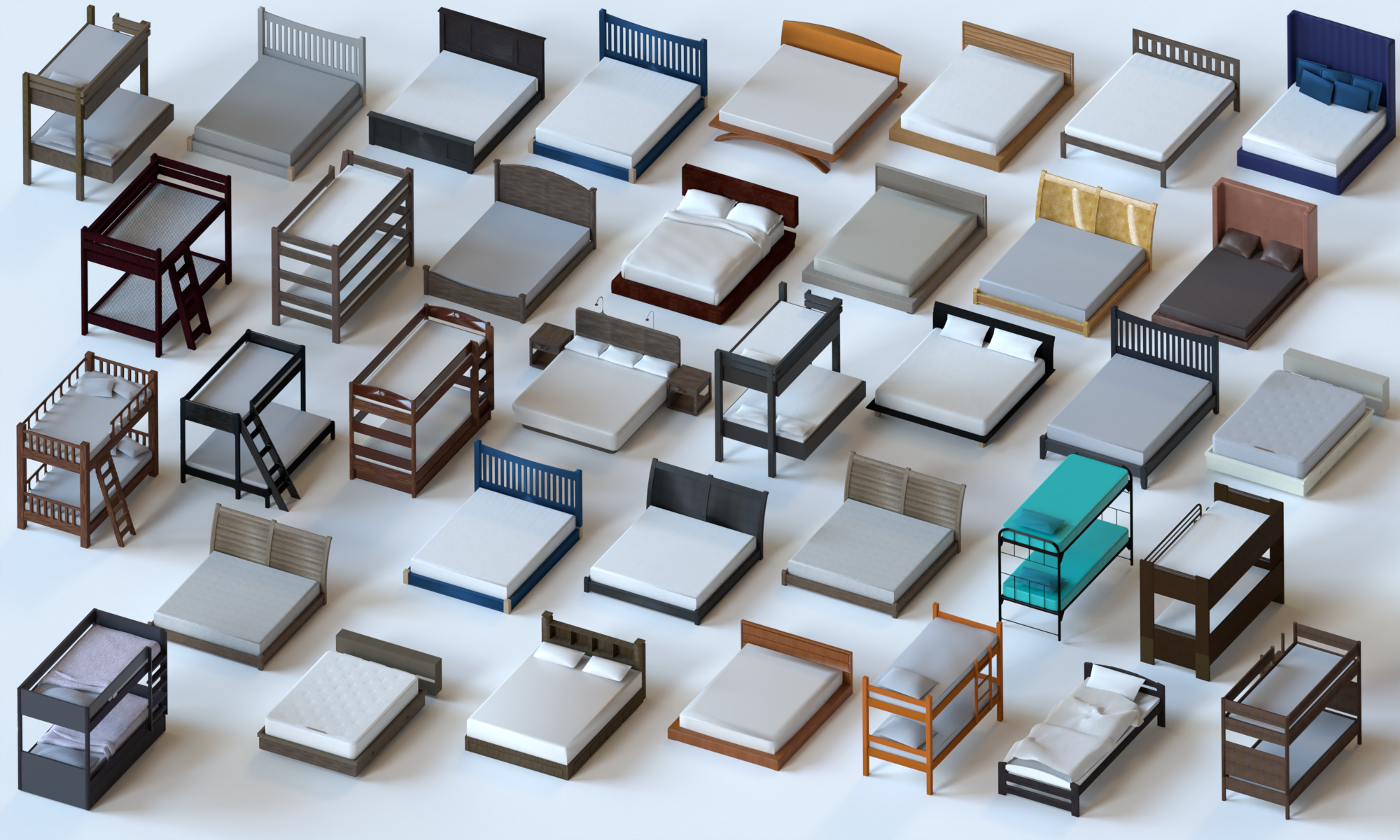}
        \caption{\change{Results for other categories of shapes to demonstrate the generality of our method.} %\rh{[updated, follow the same pattern in Fig.7]}
        }
	\label{fig:other}
\end{figure}

\change{
\subsection{Application to other categories of shapes}
%[Not sure where to put the following results. ]

To demonstrate that our method is general and can also be applied to categories other than chairs, which have been the sole focus of some of the previous work~\cite{park2018photoshape}, we also show results for other categories of shapes in Figure~\ref{fig:other}. To obtain these results, we retrain the image translation network as different categories have different semantics. However, the network is trained with the same set of hyperparameters. Moreover, we use exactly the same material predictor network that was trained on the chairs.}

\section{Discussion and Future Work}

We presented a method for material transfer from photo exemplars to 3D shapes, based on a combination of image translation and material assignment according to perceptual similarity. We showed with qualitative and quantitative evaluations that, compared to other baseline methods, our method is more robust in handling objects with diverse structures and provides shapes with materials that are closer in appearance to the provided exemplars. As a consequence, given large collections of 3D shapes and exemplars, our method can automatically create a corresponding large collection of shapes with realistic materials. In addition, given a specific 3D shape and exemplar, our method is more likely to succeed in transferring the material from the exemplar to the shape than previous works, given the improvements that we introduced with our method. 

%\rh{[need to update according to the new failure cases.]}
\paragraph{Limitations.}
Our method has certain limitations, such as the ones that cause the failure cases discussed in Figure~\ref{fig:failure}. 
%As with most machine learning approaches, incorrect correspondences can occur when the input shape or exemplar are too dissimilar from the training examples. 
%On the other hand, problems such as semantic mismatches or incorrect inference for missing parts can happen since the data used for training does not include the high-level information necessary for dealing with these cases. 
%Moreover, 
For example, our method assigns materials from a database to best match the exemplar. Thus, there could still be some discrepancies between the exemplar and the appearance of the resulting 3D shape. 
\change{Moreover, all the synthetic training images are rendered using the same illumination settings. Thus, our method may incorrectly predict a material if the photo exemplar is taken under quite different lighting.
Also, currently our material prediction method is applied on each semantic part separately without considering the global consistency of the materials combined together in a single shape, which may lead to unrealistic rendering results.}

\paragraph{Future work.}
One direction for future work is to address the limitations summarized in Figure~\ref{fig:failure}. 
%Incorrect correspondences can be remedied with the enrichment of the training sets with even more shape diversity. However, some failure cases will always exist. Thus, involving some form of minimal user input for correcting correspondences would be beneficial. 
%Since our method breaks the material assignment into two steps, user input can be involved for correcting the results of the image translation, when necessary.Moreover, semantic mismatches and incorrect material assignments for missing parts could be corrected with additional ground truth information indicating the semantic compatibility between types of chairs and shape parts.
\change{Training the networks with more images rendered under different illumination settings may improve the robustness of the method. However, the additional data may also conflict with our losses that use the perceptual similarity, since it is unclear whether perception changes are caused solely by the material itself or also by the lighting. Thus, it would be interesting to explore ways to disentangle the effects of lighting from the material prediction and assignment. 
Moreover, we believe that designing a network that predicts the materials for all the parts at once while ensuring their consistency is worth exploring.}

%\rh{[updated as we now show result on more categories.]}
%In our paper, we showed results mainly for a collection of chairs. Given that chairs have a large geometric and structural variety, our method should also be successful in handling other classes of shapes, as long as there is sufficient data to train the image translation network. Thus, it would be interesting to test our method on such categories. Also, 
Our image translation network is currently class specific. Thus, one direction for future work would be to experiment with a class-independent translation network. 
Finally, it would be interesting to combine our material assignment method with large collections of high-quality textures or texture generation methods, in order to handle exemplars with more complex textures, such as geometric patterns. The use of recent inverse texture modeling approaches would be a promising approach for exploring this research direction~\cite{hu19}.

\section{ACKNOWLEDGEMENTS}
{
We thank the anonymous reviewers for their valuable comments.
This work was supported in parts by
NSFC (61872250, U2001206, U21B2023, 62161146005),
GD Talent Plan (2019JC05X328), 
GD Natural Science Foundation (2021B1515020085),
DEGP Key Project (2018KZDXM058, 2020SFKC059),
Shenzhen Science and Technology Program (RCYX20210609103121030, RCJC20200714114435012, JCYJ20210324120213036), 
the Natural Sciences and Engineering Research Council of Canada (NSERC),
and Guangdong Laboratory of Artificial Intelligence and Digital Economy (SZ).}

\bibliographystyle{ACM-Reference-Format}
\bibliography{TMT}

\end{document}

\title{Photo-to-Shape Material Transfer for Diverse Structures\\
Supplemental Material}

\author{Ruizhen Hu}
\email{ruizhen.hu@gmail.com}
\affiliation{
 \institution{Shenzhen University}
 \country{China}
}

\author{Xiangyu Su}
\email{xiangyv.su@gmail.com}
\affiliation{
 \institution{Shenzhen University}
 \country{China}
}

\author{Xiangkai Chen}
\email{cxk19971105@gmail.com}
\affiliation{
 \institution{Shenzhen University}
 \country{China}
}

\author{Oliver van Kaick}
\email{ovankaic@gmail.com}
\affiliation{
 \institution{Carleton University}
 \country{Canada}
}

\author{Hui Huang}
\authornote{Corresponding author: Hui Huang (hhzhiyan@gmail.com).}
\email{hhzhiyan@gmail.com}
\affiliation{%
 \institution{Shenzhen University}
 \country{China}
}

%\begin{abstract}
%\input{abstract}
%\end{abstract}

\begin{CCSXML}
	<ccs2012>
	<concept>
	<concept_id>10010520.10010553.10010562</concept_id>
	<concept_desc>Computing methodologies~Computer graphics</concept_desc>
	<concept_significance>500</concept_significance>
	</concept>
	<concept>
	<concept_id>10010520.10010575.10010755</concept_id>
	<concept_desc>Computing methodologies~Shape modeling</concept_desc>
	<concept_significance>500</concept_significance>
	</concept>
	<concept>
	<concept_id>10010147.10010371.10010396.10010398</concept_id>
	<concept_desc>Computing methodologies~Mesh geometry models</concept_desc>
	<concept_significance>500</concept_significance>
	</concept>
	</ccs2012>
\end{CCSXML}

\ccsdesc[500]{Computing methodologies~Computer graphics}
\ccsdesc[500]{Computing methodologies~Shape modeling}
\ccsdesc[500]{Computing methodologies~Mesh geometry models}

%
%\keywords{}

\maketitle

\section{Datasets}
\label{sec:datasets}

\paragraph{Data collection.}
As in the work of Park \etal~\shortcite{park2018photoshape}, we use three types of datasets with our method: shape, photo, and material collections. For the photo collection, we directly use the set provided by Park \etal~\shortcite{park2018photoshape}, which consists of 40,927 chair photos. In addition, we collect 21,643 table photos and 8,787 bed photos. We explain how we set up the other two datasets as follows.
%We then explain the changes that we made for the other two datasets. 

%\paragraph{3D shape collection.}
For the shape collection, we use 3D shapes from PartNet~\cite{mo2019partnet}, where the shapes are annotated with fine-grained, instance-level, and hierarchical 3D part information. According to the analysis presented by Park \etal~\shortcite{park2018photoshape}, about $17.98\%$ of their material assignment failure cases are caused by the under-segmentation of the 3D shapes, which is one of the main causes of failures other than misalignments. Thus, the detailed semantic information provided in PartNet can help to resolve this issue. 
%As in the work of Park \etal~\shortcite{park2018photoshape}, we also focus our experiments on chairs due to the diverse structures and material combinations that appear in this class of shapes. 
The set of shapes that we use consists of 6,400 chairs with 57 semantic part labels, 7,785 tables with 82 semantic part labels, and 220 beds with 24 semantic part labels.

%\paragraph{Material collection.}
We base our material collection on the set provided by Park \etal~\shortcite{park2018photoshape}, where we have five different types of  \xy{categories} that provide a grouping of materials: leathers, fabrics, woods, metals, and plastics. However, we found that, in the material collection provided by Park \etal~\shortcite{park2018photoshape}, there exist very similar materials of the same \xy{category} or even in different \xy{category} groups.
%within the same category or even in different categories. 
Thus, we manually delete these ambiguous materials and extend the dataset by adding new, more distinctive materials, which results in a set of 600 materials in total, with 76 leathers, 197 fabrics, 160 woods, 103 metals, and 64 plastics. 

Figure~\ref{fig:material} shows the rendering of representative materials that are used for computing the perceptual similarity of materials.
%\rh{[Report the difference between our dataset and PhotoShape dataset.]}
To provide some indication to show that our material dataset has more variations, we compute the average perceptual distance for materials of the same and different \xy{categories}. Table~\ref{tab:comp_materials} shows a comparison between our dataset and the dataset of Park \etal~\shortcite{park2018photoshape}. The higher values indicate that our dataset is more diverse in the selection of materials.

\begin{figure}[!t]
	\centering
	\includegraphics[width=\linewidth]{images/material.pdf}
	\caption{
		Examples of representative materials rendered in a standard scene. These images are used for computing the perceptual similarity of materials.}
	\label{fig:material}
\end{figure}

\begin{table}[!t]%
	\newcommand{\tabincell}[2]{\begin{tabular}{@{}#1@{}}#2\end{tabular}}
    \caption{Comparison of our material collection to the collection used by Photoshape. The top line in each cell is the average perceptual distance for the given \xy{category} in our material set, while the bottom line is the perceptual distance for Photoshape's material set. Higher values indicate more diversity.} 
	\label{tab:comp_materials}
	\begin{minipage}{\columnwidth}
		\begin{center}
			\begin{tabular}{l|l|l|l|l|l}
				\hline\hline
				\textbf{\xy{Category}}& \textbf{Fabric}&  \textbf{Leather}  &   \textbf{Metal} & \textbf{Plastic} & \textbf{Wood} \\ \hline \hline 
				\tabincell{l}{Fabric} & \tabincell{l}{\textbf{42.25} \\ 35.27} \\ \hline
				
				\tabincell{l}{Leather} & \tabincell{l}{\textbf{40.24} \\ 35.27} & \tabincell{l}{\textbf{30.74} \\ 29.02} \\ \hline
				
				\tabincell{l}{Metal} & \tabincell{l}{\textbf{42.76} \\ 39.85} & \tabincell{l}{\textbf{36.26} \\ 35.93} & \tabincell{l}{\textbf{38.89} \\ 38.46} \\ \hline
				
				\tabincell{l}{Plastic} & \tabincell{l}{\textbf{49.90} \\ 43.72} & \tabincell{l}{\textbf{45.04} \\ 40.65} & \tabincell{l}{\textbf{47.94} \\ 45.29}  & \tabincell{l}{\textbf{53.95} \\ 48.42} \\ \hline
				
				\tabincell{l}{Wood} & \tabincell{l}{\textbf{41.46} \\ 36.63} & \tabincell{l}{\textbf{34.69} \\ 32.53} & \tabincell{l}{38.49 \\ \textbf{38.93}}  & \tabincell{l}{\textbf{47.67} \\ 43.68} & \tabincell{l}{\textbf{30.15} \\ 27.12} \\ \hline
			\end{tabular}
		\end{center}
	\end{minipage}
\end{table}%

\paragraph{Training data.}
Given these three datasets, we synthesize the training data for our networks. The training data is composed of (shape, image) pairs with corresponding parts and materials. To synthesize the data, we use a process similar to that described by Park \etal~\shortcite{park2018photoshape}. We first query the datasets for (shape, image) pairs where the object in the image aligns well with a projection of the 3D shape. Then, we infer the \xy{categories} for each part of the aligned objects~\cite{bell15matrec}. After that, we randomly sample a material of the corresponding \xy{category} for each part in the (shape, image) pair. 
Then, we define texture coordinates on the parts to apply the material texture.
%If the material involves the application of a texture, we also define texture coordinates on the parts. 
Similarly to Park \etal~\shortcite{park2018photoshape}, we use Blender's Smart UV projection algorithm.
%\rh{[This could be moved to the datasets section  since we need those UV maps to generate the synthetic images.]}
Finally, we render synthetic images of the aligned shape from different views.

One key difference of our data preparation is that, since we use the PartNet dataset, which has finer part segmentation, randomly assigning materials may provide fragmented and unrealistic rendering results. Thus, we group different semantic parts to obtain more realistic results. Specifically, we find that the material segmentation in ShapeNet is more natural and there is a certain correspondence between ShapeNet and PartNet shapes. Thus, we use the ShapeNet material as a prior to guide the PartNet material grouping process. In more details, we first render the same shape in PartNet and ShapeNet from the same views to obtain a semantic segmentation and material segmentation, respectively. Then, we group together the semantic parts that correspond to the same material in the material segmentation.
%, providing a more natural rendering result.
%\xy{Then, we count group operations for each semantic part, filter out less frequent merges, and select some reasonable group operations manually.} \ov{[What is a group operation?]} \rh{[I think it means how frequent different semantic parts share the same material in the default material assignment provided by ShapeNet.]} \ov{[So, maybe explain like this?]}
After that, we count how many parts were merged together, filter out less frequent merges, and select reasonable merges manually. For each shape in PartNet, we use this material group prior to group semantic labels, and assign the same material to all semantic parts within the same group with a voting process.

\paragraph{Dataset split.}
We split the shape dataset into training and test sets roughly in a 10:1 ratio. We render each shape in 5, 4, 40 different views for each chair, table, bed respectively. Finally, we get 20,085 (shape, image) training pairs and 2,000 test pairs for chairs, 28,306 training pairs and 2,832 test pairs for table, and 8,000 training pairs and 800 test pairs for bed.

\section{Network details}
\subsection{Camera pose estimation}

The camera pose estimation network is the same as the one used in~\cite{xu2019disn}. The details of the network architecture are illustrated in Table~\ref{tab:camera_pose_estimation_network}. The loss function is defined as:
\begin{equation}
\label{eq:cam}
L_{cam}  =  \frac{\sum_{p_w\in{PC_w}}{\lVert p_G - (R{p_w} + t)\rVert_2^2}}{\sum_{p_w\in{PC_w}}1}, 
\end{equation}
where $PC_w\in\mathbb{R}^{N\times3}$ is the point cloud in world space, and $N$ is the number of points in $PC_w$. For each $p_w \in{PC_w}$, $p_G$ represents the corresponding ground truth point location in the camera space, $R\in\mathbb{R}^{3\times3}$ is the predicted rotation matrix, $t\in\mathbb{R}^3$ is the predicted translation vector, and $\lVert \cdot \rVert_2^2$ is the squared $L_2$ distance.
% \usepackage{multirow}
\begin{table*}[!t]
	\caption{Camera pose estimation network structure.}
	\label{tab:camera_pose_estimation_network}
	\begin{tabular}{|c|c|c|c|c|c|c|c|c|}
		\hline
		\multicolumn{3}{|c|}{\textbf{Module}}                                                                                                                                                                                                                     & \multicolumn{3}{c|}{\textbf{Layers in the module}} & \multicolumn{3}{c|}{\textbf{Output shape(H×W×C)}} \\ \hline
		\multicolumn{3}{|c|}{\multirow{8}{*}{\begin{tabular}[c]{@{}c@{}}Feature Extraction\\ (VGG16)\end{tabular}}}                                                                                                                                      & \multicolumn{3}{c|}{Conv2d/k3s1}          & \multicolumn{3}{c|}{224×224×64}          \\ \cline{4-9} 
		\multicolumn{3}{|c|}{}                                                                                                                                                                                                                           & \multicolumn{3}{c|}{Conv2d/k3s1}          & \multicolumn{3}{c|}{112×112×128}         \\ \cline{4-9} 
		\multicolumn{3}{|c|}{}                                                                                                                                                                                                                           & \multicolumn{3}{c|}{Conv2d/k3s1}          & \multicolumn{3}{c|}{56×56×256}           \\ \cline{4-9} 
		\multicolumn{3}{|c|}{}                                                                                                                                                                                                                           & \multicolumn{3}{c|}{Conv2d/k3s1}          & \multicolumn{3}{c|}{28×28×512}           \\ \cline{4-9} 
		\multicolumn{3}{|c|}{}                                                                                                                                                                                                                           & \multicolumn{3}{c|}{Conv2d/k3s1}          & \multicolumn{3}{c|}{14×14×512}           \\ \cline{4-9} 
		\multicolumn{3}{|c|}{}                                                                                                                                                                                                                           & \multicolumn{3}{c|}{Conv2d/k7s1}          & \multicolumn{3}{c|}{1×1×4096}            \\ \cline{4-9} 
		\multicolumn{3}{|c|}{}                                                                                                                                                                                                                           & \multicolumn{3}{c|}{Conv2d/k1s1}          & \multicolumn{3}{c|}{1×1×4096}            \\ \cline{4-9} 
		\multicolumn{3}{|c|}{}                                                                                                                                                                                                                           & \multicolumn{3}{c|}{Conv2d/k1s1}          & \multicolumn{3}{c|}{1024}                \\ \hline
		\multirow{3}{*}{\begin{tabular}[c]{@{}c@{}}Scale\\ Regression\end{tabular}} & \multirow{3}{*}{\begin{tabular}[c]{@{}c@{}}Rotation\\ Regression\end{tabular}} & \multirow{3}{*}{\begin{tabular}[c]{@{}c@{}}Translation\\ Regression\end{tabular}} & FC           & FC           & FC          & 64          & 512          & 128         \\ \cline{4-9} 
		&                                                                                &                                                                                   & FC           & FC           & FC          & 32          & 256          & 64          \\ \cline{4-9} 
		&                                                                                &                                                                                   & FC           & FC           & FC          & 1           & 6            & 3           \\ \hline
	\end{tabular}
\end{table*}

\subsection{Image translation}
The image translation network is the same as the one used in~\cite{zhang20imagetrans}. The details of the network architecture are illustrated in Table~\ref{tab:image_translation_network}. The loss function is defined as:
\begin{equation}
\label{eq:trans}
L_t  =  \xy{L_{\text{col}}} + L_{\text{seg}} +  L_{\text{reg}}, 
\end{equation}

where \xy{$L_{\text{col}}$ }is the \xy{color} loss, $L_{\text{seg}}$ is the segmentation loss, and $L_{\text{reg}}$ is the regularization loss.

\subsubsection{\xy{Color} loss.}
The loss \xy{$L_{\text{col}}$} is mainly used to constrain the translated \xy{color $\hat{O}_c$} , and is defined as:
\begin{equation}
\label{eq:tex}
\xy{L_{\text{col}}} = \xy{\psi_1 L_{\text{context}}^{\text{col}}} + \psi_2 L_{\text{perc}}+  \psi_3 L_{\text{feat}} + \psi_4 L_{\text{adv}} ,
\end{equation}

Specifically,
\begin{equation}
\label{eq:tex_context}
\xy{L_{\text{context}}^{\text{col}}} = \sum_l{w_l\left[-\log\left(\frac{1}{n_l}\sum_i\max_jA^l(\phi_i^l(\xy{\hat{O}_c}), \phi_j^l(\xy{P_c}))\right)\right]} ,
\end{equation}
where $i$ and $j$ index the feature map extracted by layer $\phi^l$ of a pre-trained VGG that contains $n_l$ features, and $w_l$ controls the relative importance of different layers. We use $relu2\_2$ up to $relu5\_2$ layers of features of VGG.
\begin{equation}
\label{eq:tex_prec}
L_{\text{perc}} = {\lVert\phi_l(\xy{\hat{O}_c}) - \phi_l(\xy{O_c})\rVert_1},
\end{equation}
where we choose $\phi_l$ to be the activation after the $relu4\_2$ layer in the pre-trained VGG, since this layer mainly contains high-level semantic features.

\begin{equation}
\label{eq:feat}
L_{\text{feat}} = {\sum_l\lambda_l\lVert\phi_l(G(O_s, \xy{O^{\prime}_c})) - \phi_l(\xy{O_c})\rVert_1},
\end{equation}
where we apply random geometric distortion to \xy{$O_c$} to get the distorted image $\xy{O^{\prime}_c} = h(\xy{O_c})$, where $h$ denotes the augmentation operation like random flipping. When $O_s$ and \xy{$O^{\prime}_c$} are input, the expected output is \xy{$O_c$}. $\phi_l$ represents the feature map extracted by VGG layer $l$ and $\lambda_l$ balances the terms.

Since we use a GAN to obtain a realistic translated image \xy{$\hat{O}_c$}, the adversarial objectives of the generator $G$ and discriminator $D$ are defined as:
\begin{equation}
\label{eq:adv_G}
L_{\text{adv}}^{\text{D}} = {-\mathbb{E}[h(D(\xy{P_c}))] - \mathbb{E}[h(-D(G(O_s, \xy{P_c})))]},
\end{equation}

\begin{equation}
\label{eq:adv_D}
L_{\text{adv}}^{\text{G}} = {-\mathbb{E}[D(G(O_s, \xy{P_c}))]},
\end{equation}

\subsubsection{Segmentation loss}	
The loss $L_{\text{seg}}$  is mainly used to constrain the translated segmentation $\hat{P}_s$, and is defined as:
\begin{equation}
\label{eq:seg}
L_{\text{seg}} = \psi_5 L_{\text{pred}} + \psi_6 L_{\text{context}}^{\text{seg}},
\end{equation}
Specifically, $L_{\text{pred}}$ is the negative log likelihood loss, which is widely used in classification tasks. We use this loss to minimize pixel-wise differences between $P_s$ and $\hat{P}_s$. And $L_{\text{context}}^{\text{seg}}$ is a higher level segmentation loss which is defined as
%\begin{equation}
%\label{eq:seg_pred}
%L_{\text{pred}} = {NLL\_Loss(\hat{P}_s), P_s},
%\end{equation}
%here we use a classical classification loss to minimize the pixel-wise differences between $P_s$ and $\hat{P}_s$.
\begin{equation}
\label{eq:seg_context}
L_{\text{context}}^{\text{seg}} = {\sum_l{\lVert\omega_l(\hat{P}_s) - \omega_l(P_s)\rVert_2}},
\end{equation}
where we use another pre-trained VGG to constrain the transferred segmentation $\hat{P}_s$ to be similar to $P_s$, and we choose $\omega_l$ to be the activation after the $relu1\_2$ and $relu2\_2$ layers.

\subsubsection{Regularization loss}
The loss $L_{\text{reg}}$ is used to regularize the embedding in the shared domain of both inputs, and is defined as:
\begin{equation}
\label{eq:reg}
L_{\text{reg}} = \psi_7 L_{\text{align}} + \psi_8 L_{\text{cycle}} ,
\end{equation}
Specifically,
\begin{equation}
\label{eq:align}
L_{\text{align}} = \lVert{\mathcal{F}_{S\xrightarrow{}U}{O_s} - \mathcal{F}_{T\xrightarrow{}U}{\xy{P_c}}}\rVert_1,
\end{equation}
where $U$ is the shared domain. This loss helps to ensure that the transformed embeddings $O_u$ and $P_u$ lie in the same domain.
\begin{equation}
\label{eq:cycle}
L_{\text{cycle}} = \lVert{r_{p\xrightarrow{}o\xrightarrow{}p} - \xy{P_c}}\rVert_1,
\end{equation}
where $r_{p\xrightarrow{}o\xrightarrow{}p}$ is the forward-backward warping image.
% \usepackage{multirow}
\begin{table*}[!t]
	\caption{Image translation network structure.}
	\label{tab:image_translation_network}
	\begin{tabular}{|c|c|c|c|}
		\hline
		\textbf{Sub-network}                                                                       & \textbf{Module}                                  & \textbf{Layers in the module}   & \textbf{Output shape(H×W×C)} \\ \hline
		\multirow{9}{*}{\begin{tabular}[c]{@{}c@{}}Correspondence\\ Network\end{tabular}} & \multirow{6}{*}{Domain adaptor×2}       & Conv2d/k3s1            & 256×256×64          \\ \cline{3-4} 
		&                                         & Conv2d/k4s2            & 128×128×128         \\ \cline{3-4} 
		&                                         & Conv2d/k3s1            & 128×128×256         \\ \cline{3-4} 
		&                                         & Conv2d/k4s2            & 64×64×512           \\ \cline{3-4} 
		&                                         & Conv2d/k3s1            & 64×64×512           \\ \cline{3-4} 
		&                                         & Resblock×3/k3s1        & 64×64×512           \\ \cline{2-4} 
		& \multirow{2}{*}{Adaptive feature block} & Resblock×4             & 256×256×66          \\ \cline{3-4} 
		&                                         & Conv2d/k1s1            & 256×256×67          \\ \cline{2-4} 
		& Correspondence                          & Correlation\&warping   & 256×256×68          \\ \hline
		\multirow{8}{*}{\begin{tabular}[c]{@{}c@{}}Translation\\ Network\end{tabular}}    & \multirow{3}{*}{Style encoder×7}        & Bilinear interpolation & h×w×3               \\ \cline{3-4} 
		&                                         & Conv2d/k3s1            & h×w×128             \\ \cline{3-4} 
		&                                         & Conv2d/k3s1            & h×w×c               \\ \cline{2-4} 
		& \multirow{5}{*}{Generator}              & Conv2d/k3s1            & 8×8×1024            \\ \cline{3-4} 
		&                                         & Resblock×5             & 128×128×256         \\ \cline{3-4} 
		&                                         & Nonlocal               & 128×128×256         \\ \cline{3-4} 
		&                                         & Resblock×2             & 256×256×64          \\ \cline{3-4} 
		&                                         & Conv2d/k3s1            & 256×256×3           \\ \hline
	\end{tabular}
\end{table*}

\subsection{Material prediction}
The architecture of the material prediction network and loss function are described in the main paper in Section~4.2. We provide the details here again for completeness.
We use a Resnet-34~\cite{he2016deep} pre-trained on ImageNet~\cite{deng2009imagenet} as our backbone network, but remove the last layer and add a FC layer instead to embed the input images into a $128$-D feature space. We also add two FC layers after that to predict the \xy{category} and material labels, respectively. The details of the network architecture are illustrated in~\ref{tab:material_prediction_network}. The loss function for training our material prediction network is defined as follows:
\begin{equation}
\label{eq:matpred}
L_p = L_{\text{metric}} +  L_{\text{class}} ,
\end{equation}
where $L_{\text{metric}}$ is the metric learning loss to ensure the embedded features reflect the perceptual distance, and $L_{\text{class}} $ is the loss defined on the final material assignment to make sure that it is similar to the ground truth.

\subsubsection{Metric learning loss.} 
The metric learning loss $L_{\text{metric}}$ is adapted from the work of Lagunas et al.~\shortcite{lagunas2019similarity}, and is composed of two losses defined on material triplets, where each triplet given to the network consists of a reference image $r$ with material $m_r$, one positive example $a$ with material $m_a$, and one negative example $b$ with material $m_b$. 
The metric learning loss is defined as:
\begin{equation}
\label{eq:matpred}
L_{\text{metric}} =\alpha_1 L_{\text{tri}} + \alpha_2 L_{\text{sim}},
\end{equation}
where $L_{\text{tri}}$ is the triplet loss that seeks to bring similar materials $r$ and $a$ closer together in the feature space and repel dissimilar materials $b$, and $L_{\text{sim}}$ is the similarity loss that further maximizes the log-likelihood of the model choosing $a$ to be closer to $r$ than $b$.

The triplet loss is defined as follows~\cite{lagunas2019similarity}:
\begin{equation}
L_{\text{tri}}(r, a, b) = \frac{1}{|B^A|} \sum_{(r,a,b) \in B^A} \left[ \lVert f_r - f_a\rVert_2^2  - \lVert f_r - f_b\rVert_2^2  + \mu \right]_{+},
\end{equation}
where $[x]_{+} = \max(0,x)$, $f_x$ is the feature vector of $x$, $\mu$ is the margin which specifies how much we would like to separate the samples in the feature space, and $B^A$ is the set of triplets that take part in the training.
%\ov{[Should we change it to $m_r$ instead of $m^r$ for consistency with the notation above? Or are these single labels rather than vectors? Then, we can clarify this.]} \rh{[Right, let's use $m_r$ to be more consistent.]}

The similarity loss is defined as follows~\cite{lagunas2019similarity}:
\begin{equation}
L_{\text{sim}}  = -\frac{1}{|B^A|} \sum_{(r,a,b) \in B^A} \log \frac{s_{ra}}{s_{ra} + s_{rb}},
\end{equation}
where $s_{ra} = \dfrac{1}{1+\lVert f_r - f_a\rVert_2^2}$ and $s_{rb} = \dfrac{1}{1+\lVert f_r - f_b\rVert_2^2}$.
\\
\\
To construct the training triplets $B^A$, we first generate a set of pre-sampled material triplets $A^M$. To generate $A^M$, we randomly sample a reference material $m_r$, a random sampled positive material $m_a$ which has the same \xy{category} as $m_r$, and a negative material $m_b$ which has a different \xy{category} than $m_r$. The material $m_b$ is sampled so that it has a larger perceptual distance to $m_r$ than $m_a$, i.e., $D(m_r,m_b) > D(m_r,m_a)$, but also has a distance smaller than the distances to all other materials with all different \xy{category}, i.e., $D(m_r,m_b) < D(m_r,m_x)$, where $m_x$ is any material with a \xy{category} other than the \xy{category} of $m_r$. %and $m^b$.
Then  $B^A$ is defined as:
\begin{equation}
B^A = \left\{(r,a,b) \mid (m_r, m_a, m_b) \in A^M  \wedge (r, a, b)\in B \right\},
\end{equation}
where $B$ is the current training batch. Thus, $B^A$ is the set of all triplets of images in $B$ whose corresponding material labels appear as triplets in $A^M$.

	\subsubsection{Classification loss.} 
	The classification loss $ L_{\text{class}}$ for training our material prediction network is defined as follows:
	\begin{equation}
	\label{eq:matpred}
	L_{\text{class}} = \alpha_3 \xy{L_{\text{cat}}} + \alpha_4 L_{\text{mat}} + \alpha_5 L_{\text{dis}},
	\end{equation}
	where \xy{$L_{\text{cat}}$} and $L_{\text{mat}}$ are the cross entropy losses defined by Park et al.~\shortcite{park2018photoshape} for \xy{category} and material classification, and $L_{\text{dis}}$ is the distance loss to minimize the perceptual distance between the predicted material and the ground truth material:
	\begin{equation}
	\label{eq:dist}
	L_{\text{dis}}(m_p, m_{gt}) = m_p^{T} D_{\text{idx}(m_{gt})},
	\end{equation}
	where $m_p$ and $m_{gt}$ are the predicted and ground truth material labels, represented as $n$-dimensional column vectors where $n$ is the size of the material dataset. Specifically, these vectors are one-hot vectors for the ground truth labels, and probability vectors for predicted labels.
	%\ov{[Are these one-hot vectors then?]} \rh{[Yes for ground truth label. But for the predicted label it would just be a |M|-D vector indicating the probability of each label. From the previous notation, then |M| will just be $n$, right? ]}
	$D_i$ represents the $i$-th column of the matrix $D$, $\text{idx}(m_{gt})$ is the index of the ground truth material and thus $D_{\text{idx}(m_{gt})}$ encodes the perceptual distance of $m_{gt}$ to all other materials. %\rh{[Should we use different notations? Since $x$ and $y$ have been used in the image translation network. Maybe use $m_p$ and $m_{gt}$?]}
	%\ov{[Added "column" above and changed the notation a bit for the formula to make sense and L be a scalar. Is this fine?]} \rh{[Yes. Maybe we can use $D(:,m_{gt})$ to represent the corresponding column and simplify the notation since $m_{gt}$ is a one-hot vector.]}\ov{[I thought it's better to use proper mathematical notation, even though it's a bit more complicated, as some reviewer may complain]} 

% \usepackage{multirow}
\begin{table}[!t]
	\caption{Material prediction network structure.}
	\label{tab:material_prediction_network}
	\begin{tabular}{|c|c|c|c|c|c|}
		\hline
		\multicolumn{2}{|c|}{\textbf{Module}}                                                                                                                            & \multicolumn{2}{c|}{\textbf{Layers in the module}}                                            & \multicolumn{2}{c|}{\textbf{Output shape}} \\ \hline
		\multicolumn{2}{|c|}{\begin{tabular}[c]{@{}c@{}}Feature Extraction\\ (ResNet-34)\end{tabular}}                                                                   & \multicolumn{2}{c|}{\begin{tabular}[c]{@{}c@{}}ResNet-34 \\ before\\ classifier\end{tabular}} & \multicolumn{2}{c|}{512}                          \\ \hline
		\multicolumn{2}{|c|}{Feature Mapping}                                                                                                                            & \multicolumn{2}{c|}{FC}                                                                       & \multicolumn{2}{c|}{128}                          \\ \hline
		\multirow{3}{*}{\begin{tabular}[c]{@{}c@{}}Material\\ Classifier\end{tabular}} & \multirow{3}{*}{\begin{tabular}[c]{@{}c@{}}\xy{Category}\\ Classifier\end{tabular}} & FC                                             & FC                                           & 512                     & 512                     \\ \cline{3-6} 
		&                                                                                 & FC                                             & FC                                           & 512                     & 512                     \\ \cline{3-6} 
		&                                                                                 & FC                                             & FC                                           & 600                     & 5                       \\ \hline
	\end{tabular}
\end{table}

\subsection{Training details}
We pre-train the image translation network with an Nvidia RTX3090 GPU under Ubuntu 20.04.2 and Pytorch 1.7.0 for 50 epochs with parameters set as $\{\psi_1, \dots, \psi_8\}$ = $\{$ 1, 0.01, 10, 10, 100, 2, 1, 1 $\}$ \xy{using Adam optimizer with initial learning rate 2e-4 and batch size 3}, which takes about 96 hours. We pre-train the material prediction network for 40 epochs for metric learning with only the metric learning loss and $\alpha_1=\alpha_2=1$ \xy{using SGD optimizer with initial learning rate 1e-3, momentum 0.9 and batch size 180}, which takes about 20 hours. Then, we continue to train the classification layers for 180 epochs with $\alpha_3=0.5$, $\alpha_4=1$ and $\alpha_5=10$ \xy{using Adam optimizer with initial learning rate 5e-4 and batch size 180}, which takes about 30 hours. During the fine-tuning, we train the material prediction network for 10 epochs with $\alpha_1=\alpha_2=0$, $\alpha_3=0.5$, $\alpha_4=1$, and $\alpha_5=10$, which takes about 10 hours.

\section{User study}
\begin{figure}[!t]
	\centering
	\includegraphics[width=\linewidth]{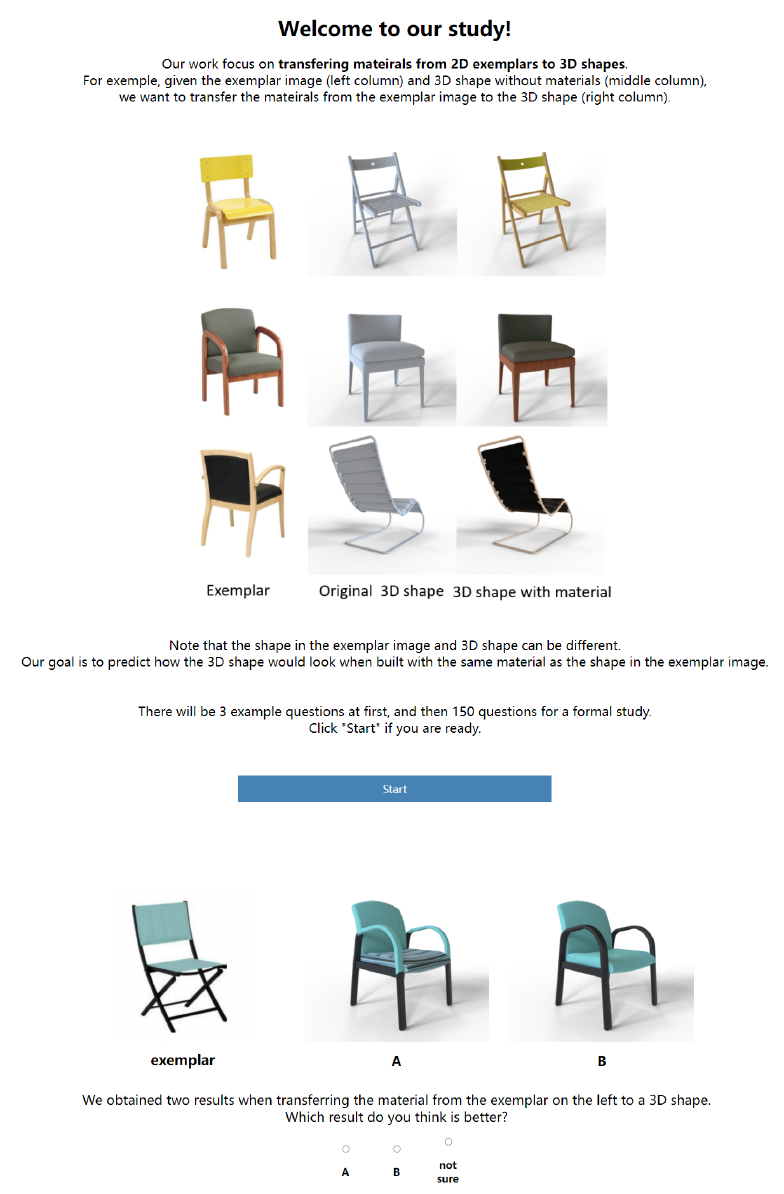}
	\caption{
		Interface used in our user study. Top: initial screen
                with instructions. Bottom: example screen with a
                question.}
	\label{fig:userstudy}
\end{figure}

The specific wording used in the instructions of our user study is ``Our work focuses on transferring materials from images to 3D shapes. For example, given the exemplar image (left column) and 3D shape without materials (middle column), we want to transfer the materials from the exemplar image to the 3D shape (right column). Note that the shape in the exemplar image and 3D shape can be different. Our goal is to predict how the 3D shape would look when built with the same material as the shape in the exemplar image.''. 

The wording used in the questions is ``We obtained two results when transferring the material from the exemplar on the left to a 3D shape. Which result do you think is better?''.

Figure~\ref{fig:userstudy} shows example screens of the interface used for
the user study.

\section{Application to other categories of shapes}
%[Not sure where to put the following results. ]
\begin{figure}[!t]
	\centering
	\includegraphics[width=\linewidth]{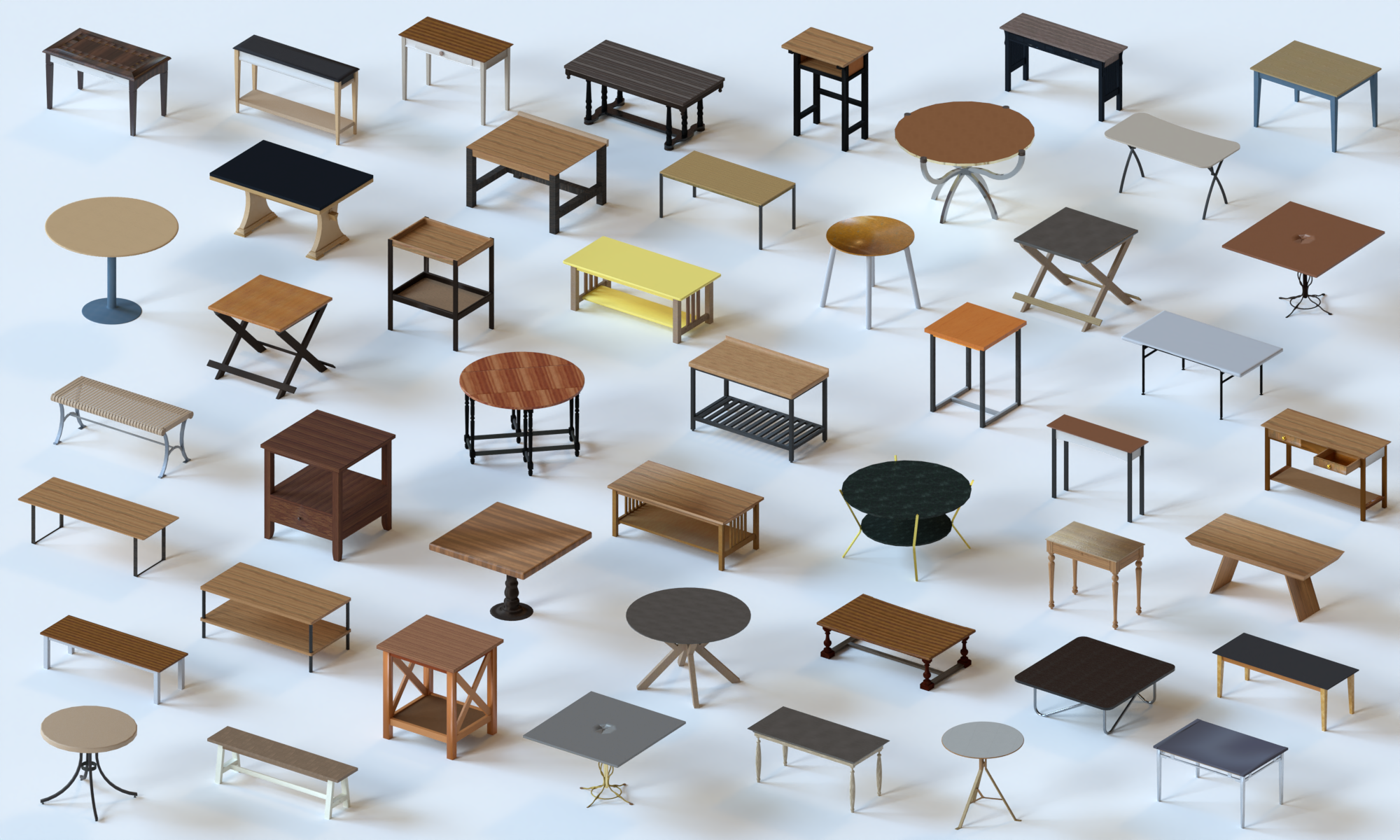}
	\includegraphics[width=\linewidth]{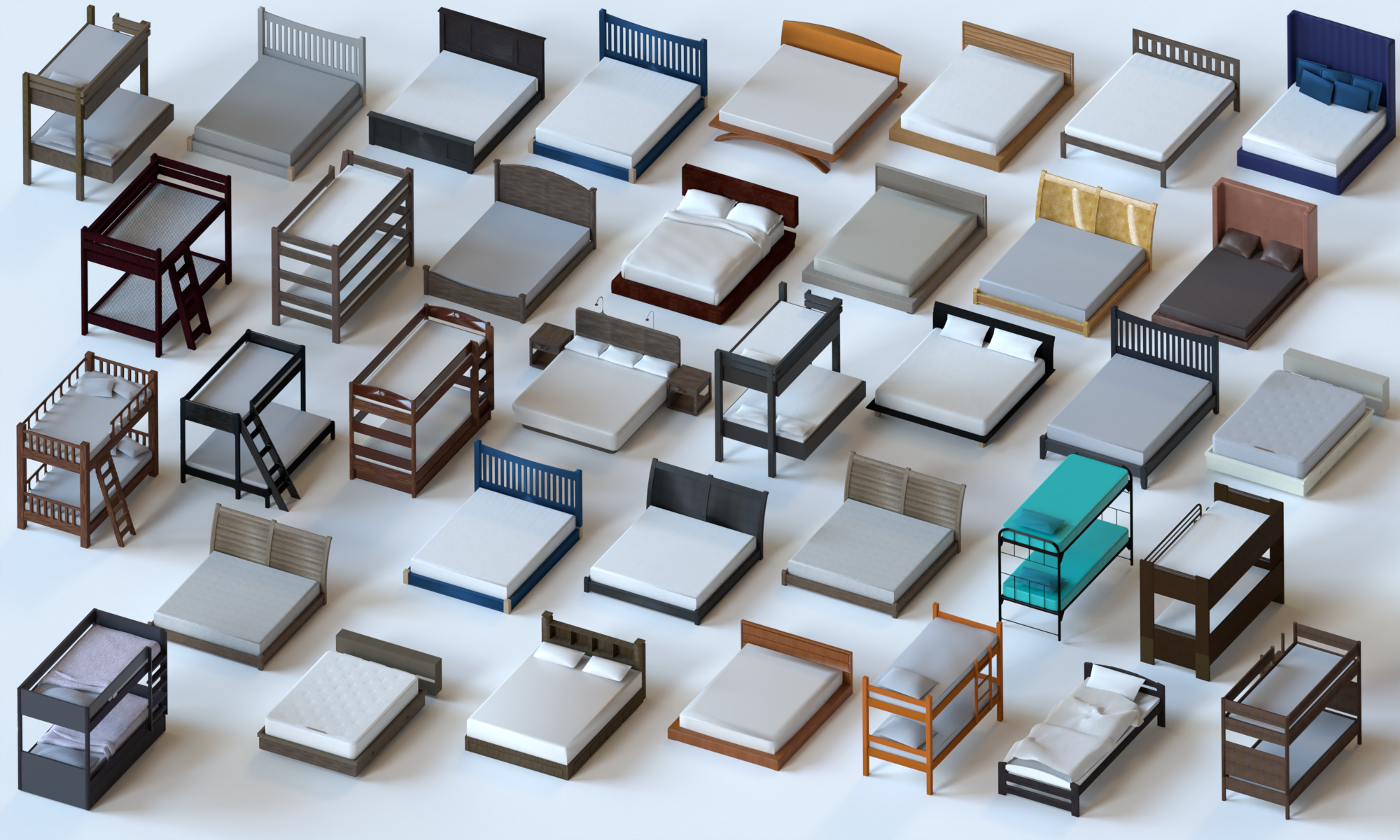}
        \caption{\change{Results for other categories of shapes to demonstrate the generality of our method.} %\rh{[updated, follow the same pattern in Fig.7]}
        }
	\label{fig:other}
\end{figure}

To demonstrate that our method is general and can also be applied to categories other than chairs, which have been the sole focus of some of the previous work~\cite{park2018photoshape}, we also show results for beds and tables in Figure~\ref{fig:other}. To obtain these results, we retrain the image translation network as different categories have different semantics. However, the network is trained with the same set of hyperparameters. Moreover, we use exactly the same material predictor network that was trained on the chairs. 

Figure~\ref{fig:other} shows a sample of results of our method on beds (row 1-3) and tables (row 4-6), where we show the input exemplar and segmentation on the left, translated segmentation and \xy{color} images in the middle, and final material transfer result on the right. We can see that the method is able to handle beds and tables with different structure and geometry. For example, the method successfully transfers the material from a normal to a bunk bed (row 3), from a square to round table (row 5), exchanges materials between table legs with different topologies (row 4 and 6). The method is also successful in transferring materials to bunk bed with ladder even though the exemplar do not have this part (row 3).

\bibliographystyle{ACM-Reference-Format}
\bibliography{TMT}